\documentclass[10pt]{article}
\usepackage{sao2}
\usepackage{psfig,epsf}

\issue{2009, 64, 272--279}

\def\apj{Astrophys. J}

\def\apjs{Astrophys. J. Supp.}
\def\ijmpd{Int. J. Mod. Phys. D}

\def\prd{Phys.\ Rev.\ D}

\def\sqd{\hbox{\rlap{$\sqcap$}$\sqcup$}$^\circ$}

\begin{document}

\markboth{O.V.~Verkhodanov, M.L.~Khabibullina, E.K.~Majorova}
       {TESSELLATED MAPPING OF COSMIC BACKGROUND RADIATION CORRELATIONS}
\title{Tessellated Mapping of Cosmic Background Radiation Correlations and Source Distributions}
\author{O.V.~Verkhodanov\inst{a}
\and M.L.~Khabibullina\inst{a}
\and E.K.~Majorova\inst{a}
}
\institute{
\saoname
}
\date{13 October, 2008}{1 December, 2008}
\maketitle

\begin{abstract}
We offer a method of correlations mapping on the full celestial
sphere that allows to check the quality of reconstructed maps,
their non-Gaussianity and conduct experiments in various frequency
ranges. The method was evaluated on the WMAP data, both on the
reconstructed maps and foreground components, and on the NRAO VLA
Sky Survey (NVSS) data. We detected a significant shift in the
correlation data of the dust component, which can be
preconditioned by a more complex dust model than the one currently
in use for component separation. While studying the NVSS
correlation data, we demonstrated that the statistics of the
coinciding spots in the microwave background and in the NVSS
survey corresponds to the one expected in the $\Lambda$CDM model.
This can testify for a chance coincidence of the spots in the NVSS
and WMAP data in the CMB Cold Spot region. Our method is
software-implemented in the GLESP package.
%
\mbox{\hspace{1cm}}
\end{abstract}

\section{INTRODUCTION}
Full celestial sphere surveys in different frequency ranges have
led to a major advance in the research of global properties of the
world around us. Such important matters as determining the
cosmological parameters of the Universe in its early stages, the
formation of the large-scale structure of the Universe and the
phenomenon of dark energy are among them. The observations of the
cosmic microwave background radiation (CMB) by the Wilkinson
Microwave Anisotropy Probe\footnote{\tt
http://lambda.gsfc.nasa.gov} (WMAP)
\cite{wmapresults:Verkhodanov_n,wmap5ytem:Verkhodanov_n,wmap5ycos:Verkhodanov_n},
were revolutionary in modern cosmology. The data were recorded in
five bands: 23\,GHz (the K-band), 33\,GHz (the Ka-band), 41\,GHz
\mbox{(the Q-band),} 61\,GHz (the V-band) and 94\,GHz \mbox{(the-W
band)} with the measurements of intensity and polarization. The
result of the analysis of these data were the CMB maps of
anisotropy and polarization, the maps of foreground components
(synchrotron and free-free emission, dust radiation), their power
spectrum were as well deduced. For the harmonics of not very high
order \mbox{($\ell<100$)}, a map of CMB anisotropy distribution is
listed, reconstructed from the multifrequency observations of
foreground components implementing the Internal Linear Combination
(ILC) method \cite{wmapresults:Verkhodanov_n}.

In present-day full sphere observations, the infrared
\cite{mass2a:Verkhodanov_n}, radio \cite{nvss:Verkhodanov_n} and
optical \cite{sdss:Verkhodanov_n} wavelength ranges survey maps
are actively used along with the microwave background data. With
the aid of correlation between the CMB and radio emission, one can
study both the contribution of radio sources into the common
background \cite{cmb_rs:Verkhodanov_n,vo_par_book:Verkhodanov_n},
and the correlation properties of the radio source and CMB
distribution occurring both on the large scale ($\theta>2\degr$),
like the Sachs-Wolfe effect \cite{swe:Verkhodanov_n}, and on the
small scale (mainly $\theta<4\arcmin$) \cite{sz:Verkhodanov_n},
like the Sunyaev-Zel'dovich effect. Correlations of infrared and
optical surveys (specifically with known red shifts) with CMB data
allow us to study the formation of a large-scale structure.

Another important application of correlation methods are studies
of CMB maps quality and their cleaning level from other kinds of
radiation. For example, the presence of correlation between the
CMB and foreground components
\cite{ndv03:Verkhodanov_n,ndv04:Verkhodanov_n} carried in by the
Galaxy, points to a residual contribution of the interfering
emission in the procedure of components separation. Therewith, the
presence of correlated components in a number of multipole
frequency ranges leads to a change in the CMB signal statistics,
that manifests itself through non-Gaussianity
\cite{nong:Verkhodanov_n}. That, in its turn, makes the analysis
of the maps of the studied signal and their power spectrum more
complex. Quite a significant number of papers has been dedicated
to the studies of statistic properties of the signal on the ILC
map and to the discussion of its non-Gaussianity. These works were
conducted using different methods, like the phase analysis
\mbox{method
\cite{ndv04:Verkhodanov_n,nong:Verkhodanov_n,coles04:Verkhodanov_n},}
Maxwell's multipole vectors method
\cite{copi:Verkhodanov_n}, wavelet analysis
\cite{vielva04:Verkhodanov_n,mw_wavel:Verkhodanov_n,cruz05:Verkhodanov_n}
and Minkowski functionals
\cite{eriksenmf:Verkhodanov_n,park3y:Verkhodanov_n}. In the
non-Gaussianity studies, the correlation method allows separating
a non-Gaussianity of a certain type, e.g. the one preconditioned
by systematization during data processing.

In this paper we generalize the approach of map correlation in
various angular scales, earlier described in
\cite{rzf_cmb:Verkhodanov_n,nvss_cmb:Verkhodanov_n}. We offer a
method of tessellated mapping and maps correlation visualisation
on the full sphere for different angular scales of correlation
search. We use this approach for the construction and study of the
ILC and WMAP5 foreground components correlation maps, where the
WMAP data are from the fifth year of observations
\cite{wmap5ytem:Verkhodanov_n}; as well as for the correlation
maps of ILC and NRAO VLA Sky Survey (NVSS) radio sources
\cite{nvss:Verkhodanov_n}.

\section{THE METHOD OF CORRELATIONS MAPPING}

The correlation of two maps on the sphere is described by a
correlation coefficient for a multipole $\ell$ as follows
\begin{equation}
K(\ell) = \frac{1}{2}
\frac{\sum\limits_{m=-\ell}^l t_{\ell m}s^*_{\ell m} + t^*_{\ell m}s_{\ell m}}
       {(\sum\limits_{m=-\ell}^l |t_{\ell m}|^2
     \sum\limits_{m=-\ell}^\ell |s_{\ell m}|^2)^{1/2}}\,,
\end{equation}
where $t_{\ell m}$ and $s_{\ell m}$ are the variations of CMB
temperature and some other signal in a harmonic representation,
* is a symbol of conjugation, $\ell$ and $m$ are the numbers of spherical harmonics
(of the multipole) and its modes in a harmonic decomposition of
the signal on the sphere:
 \begin{eqnarray}
S(\theta,\phi)=  \quad \quad \quad \quad \quad \quad \quad \quad \nonumber \\ \sum_\ell^{\ell_{max}}\sum_{m=1}^\ell\left(a_{\ell,m}
Y_{\ell,m}(\theta,\phi)+a_{\ell,-m}Y_{\ell,-m}(\theta,\phi)\right)\,,
\end{eqnarray}
where $Y_{\ell,m}$ are the spherical functions, $(\theta,\phi)$
are the spherical coordinates, $a_{\ell,m}$ are the coefficients
of spherical harmonics satisfying the following condition
\begin{eqnarray}
Y_{\ell,-m}(\theta,\phi)=(-1)^mY^*_{\ell,m}(\theta,\phi),\nonumber \\ \quad
a_{\ell,m}=(-1)^m a^{*}_{\ell,-m}\,.\quad \quad \quad
\end{eqnarray}
The value of the $K(\ell)$ coefficient allows to check the
harmonics correlation on the sphere, i.e. to compare the
properties of maps in a given angular scale. However, while
looking for the correlated regions which do not recur in other
regions of the sphere, this approach brushes out such single areas
while common averaging over the sphere within one harmonic. In
this case it is practically impossible to separate the correlated
regions.

\begin{figure*}[tbp]
\centerline{
\vbox{
\hbox{
\psfig{figure=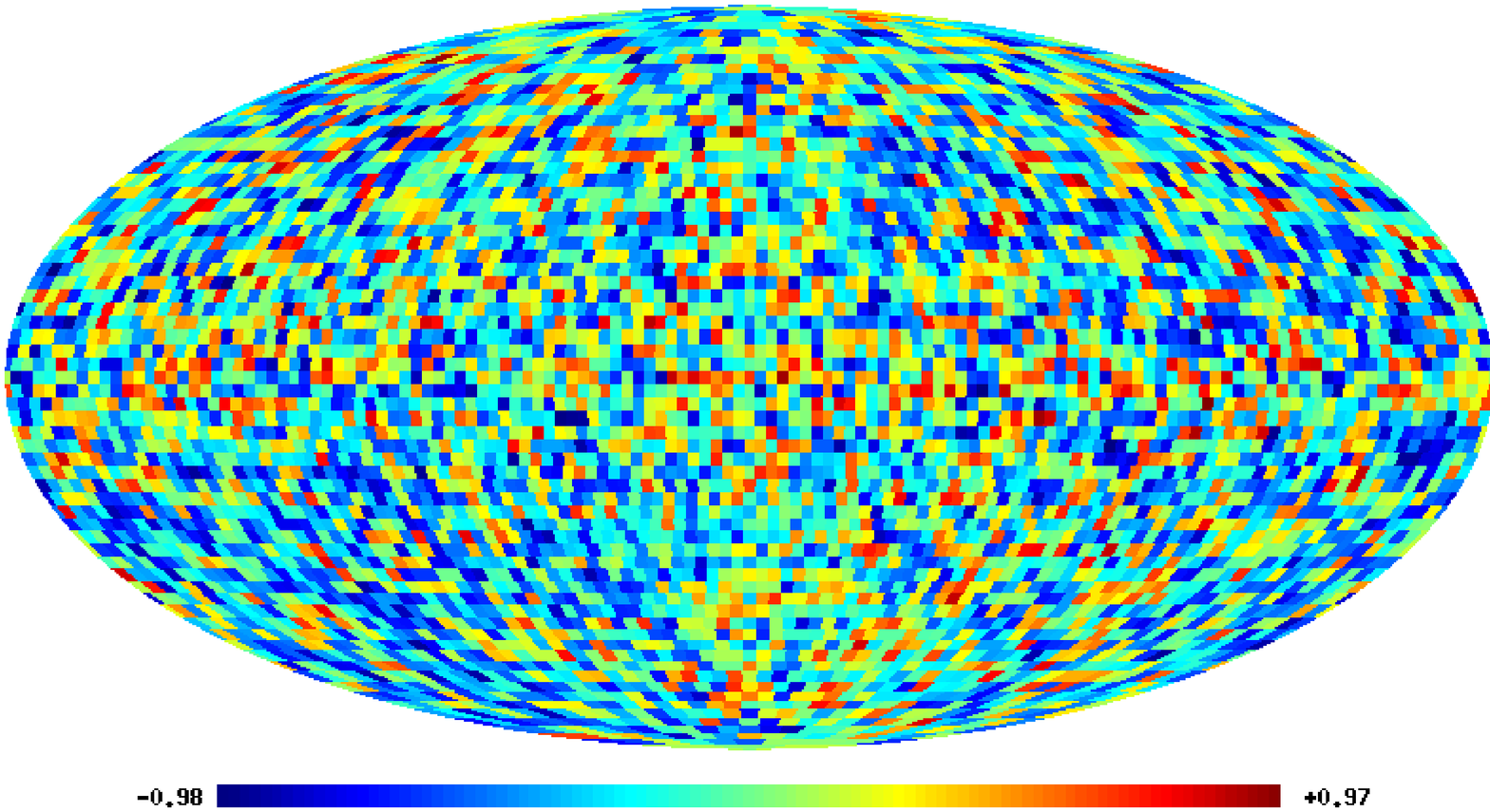,width=5cm,angle=-90}
\psfig{figure=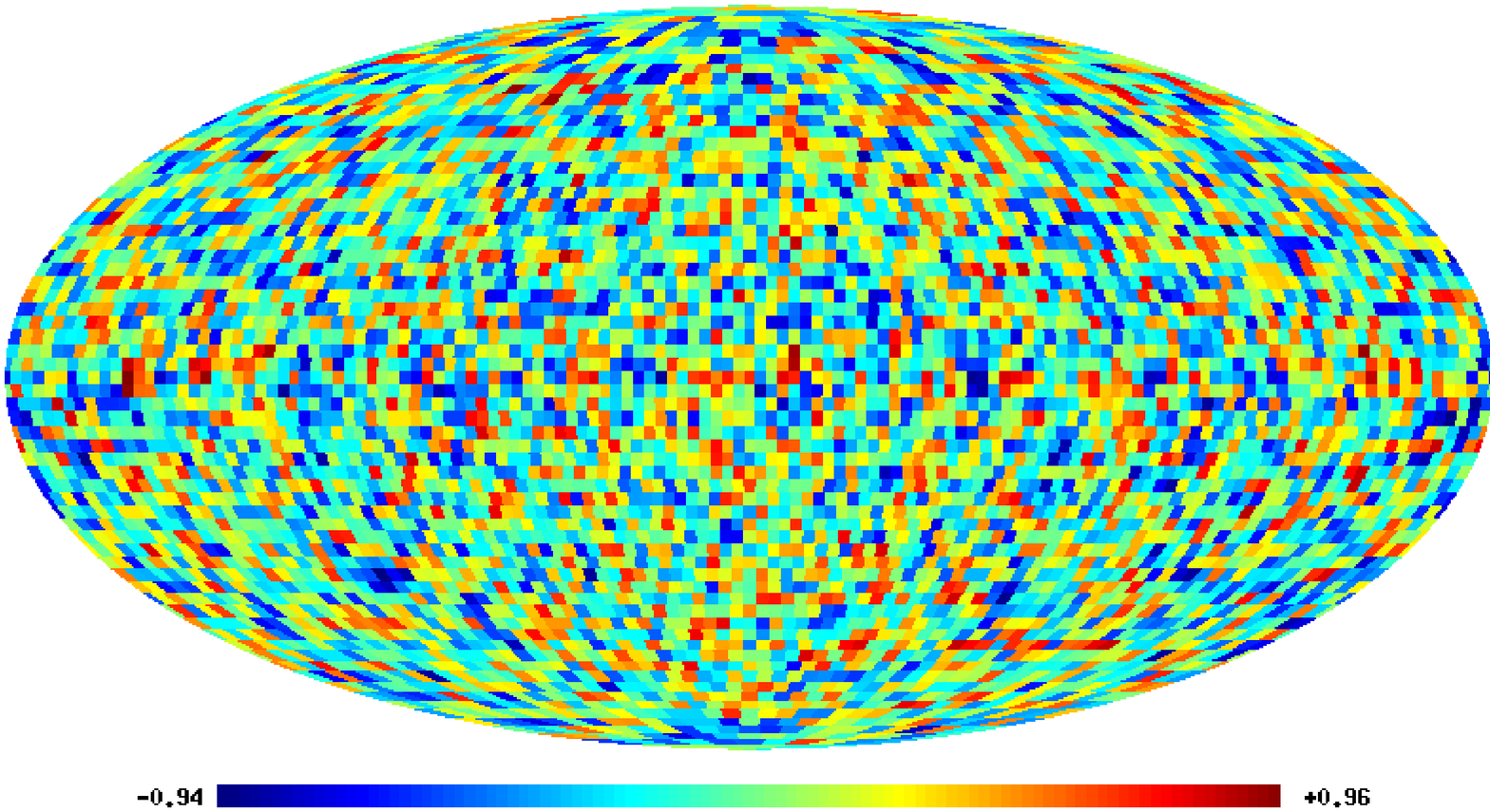,width=5cm,angle=-90}
\psfig{figure=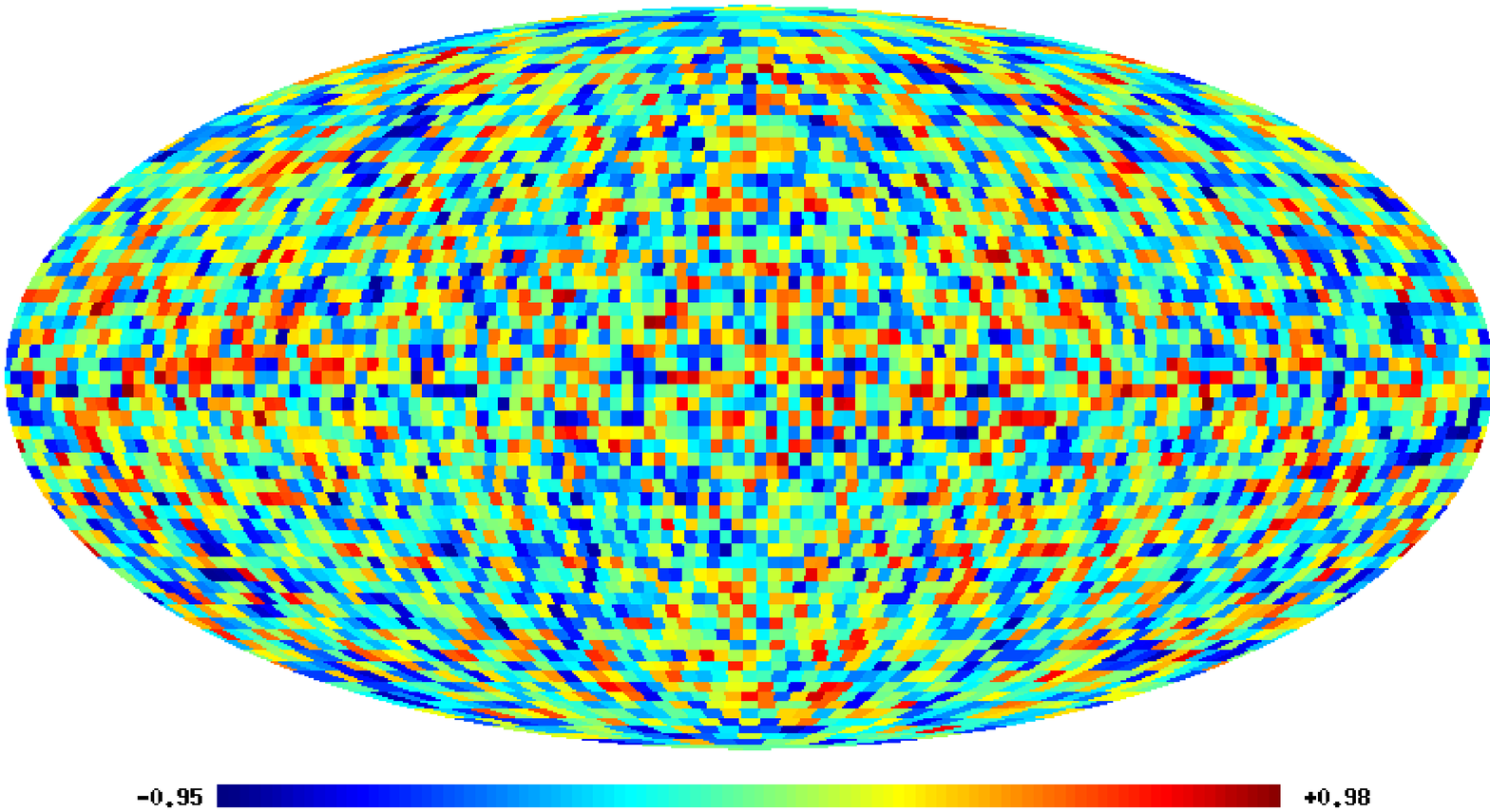,width=5cm,angle=-90}
}
\hbox{
\psfig{figure=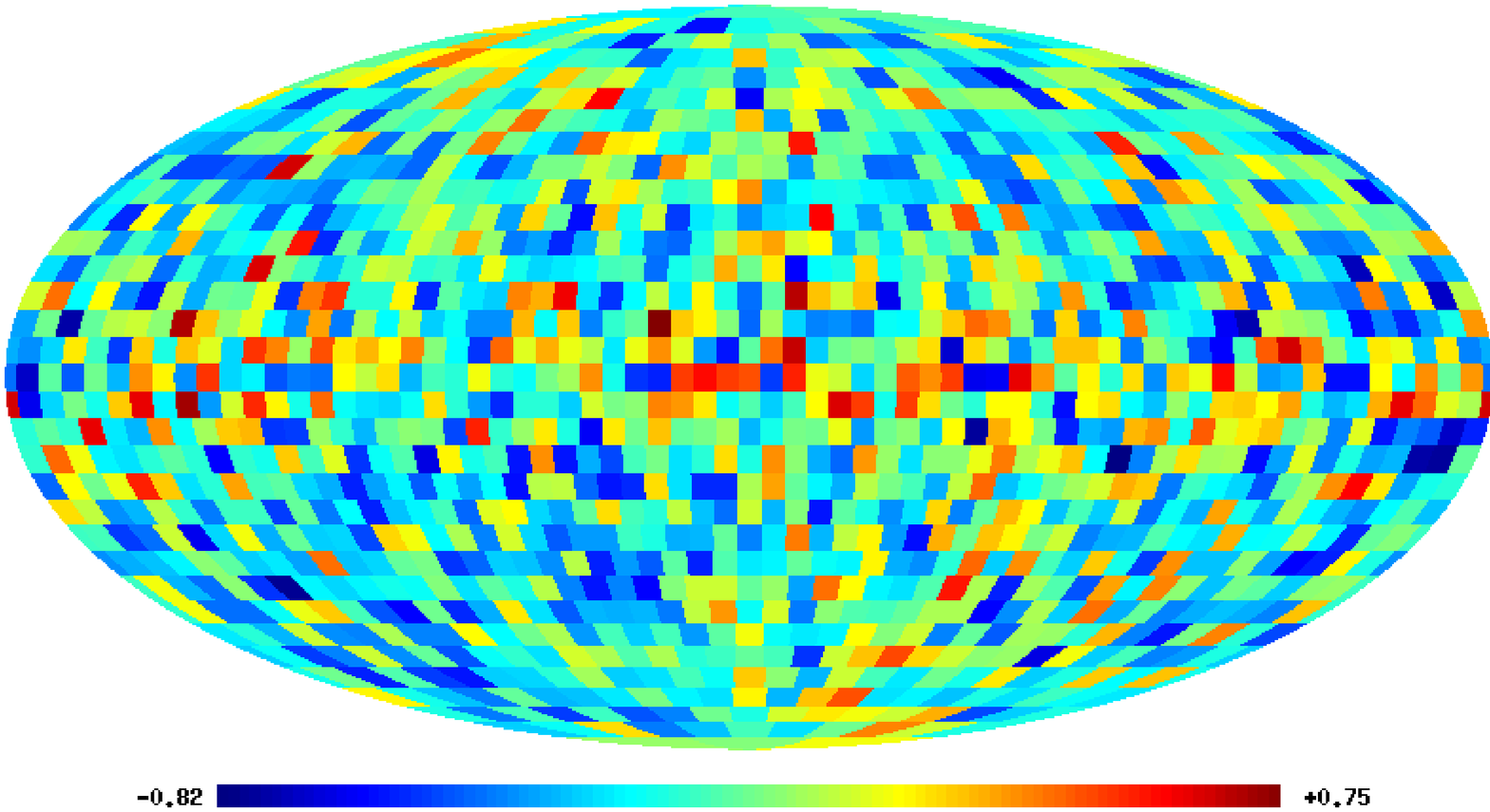,width=5cm,angle=-90}
\psfig{figure=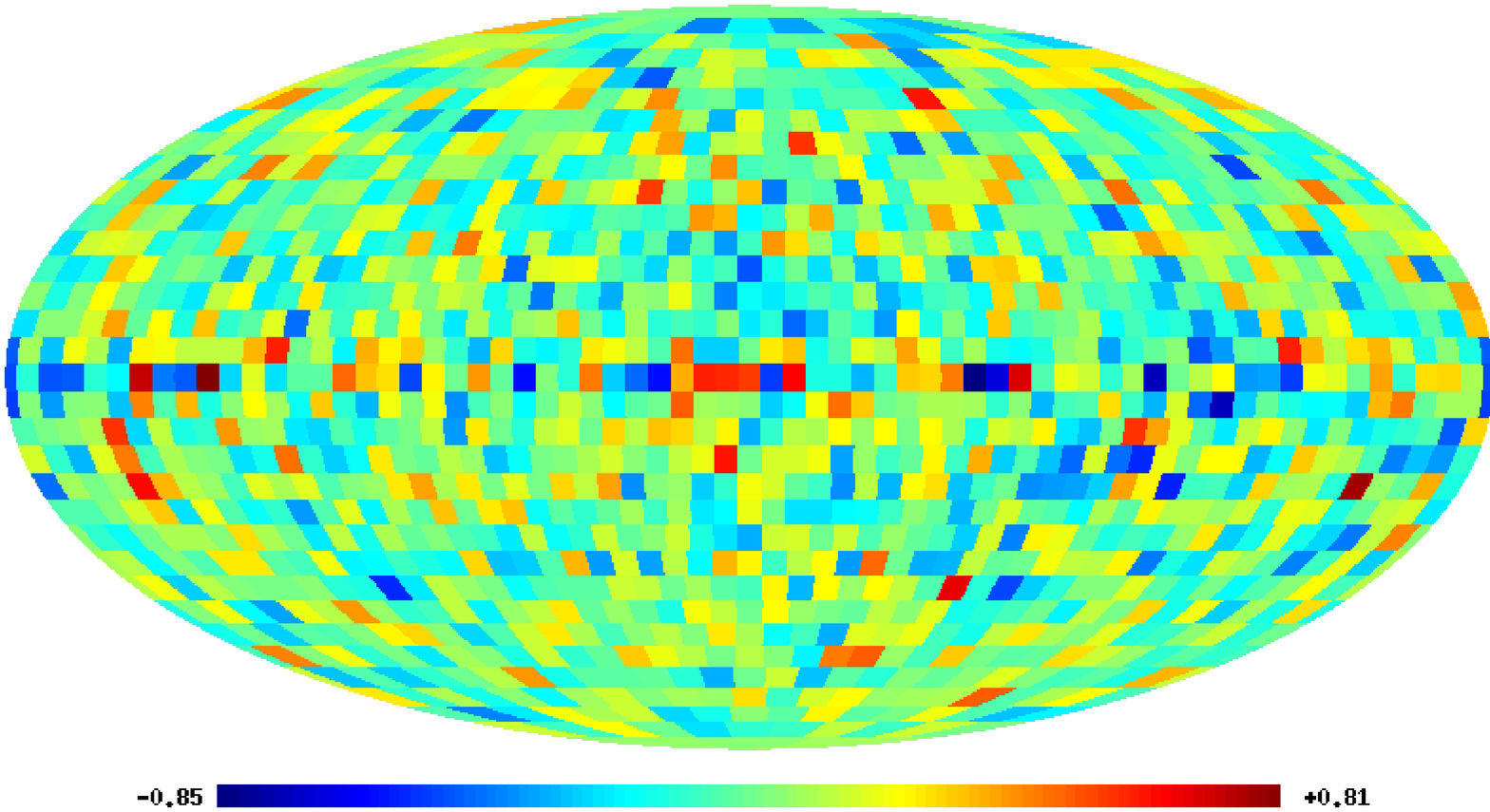,width=5cm,angle=-90}
\psfig{figure=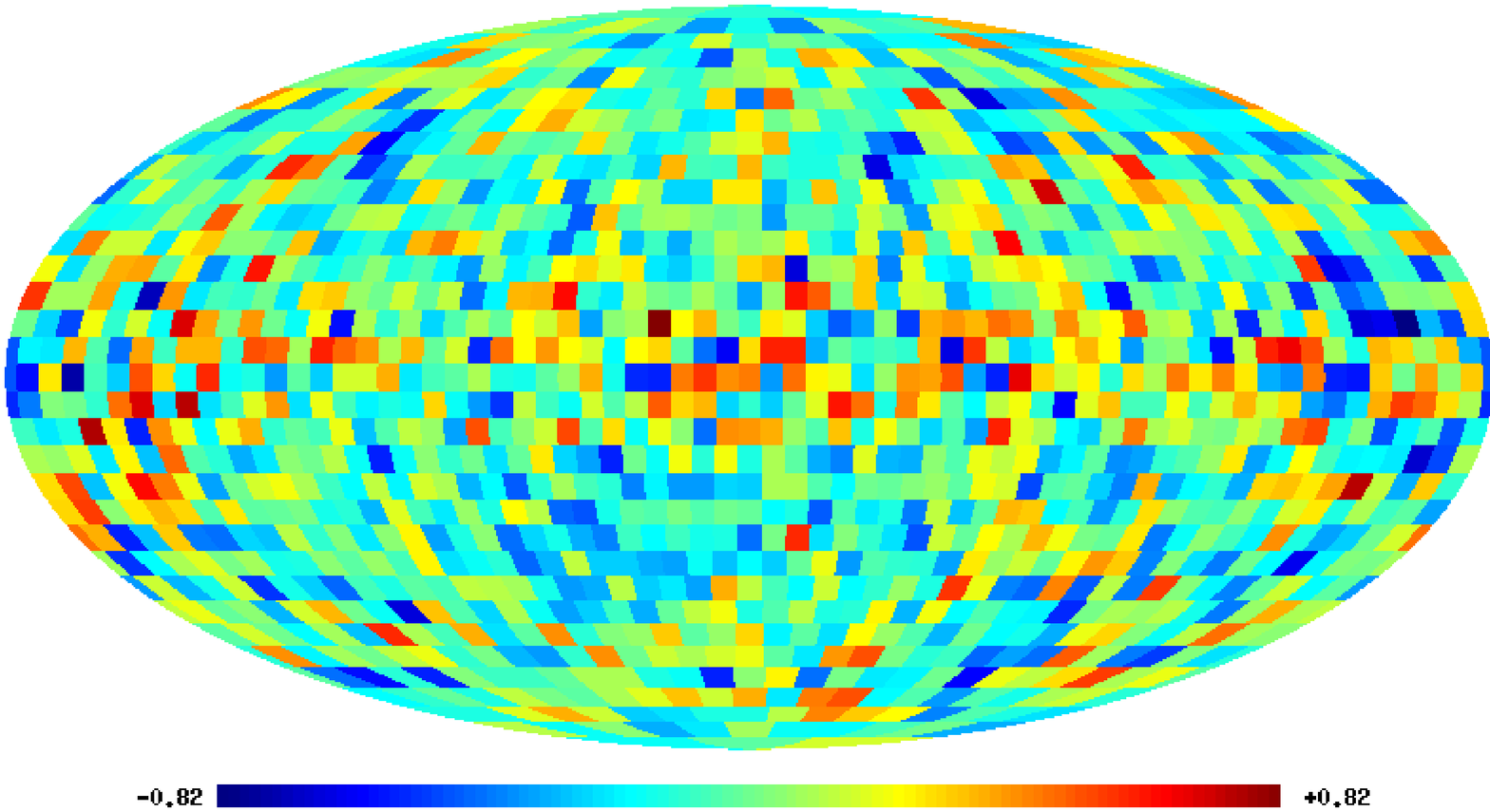,width=5cm,angle=-90}
}
\hbox{
\psfig{figure=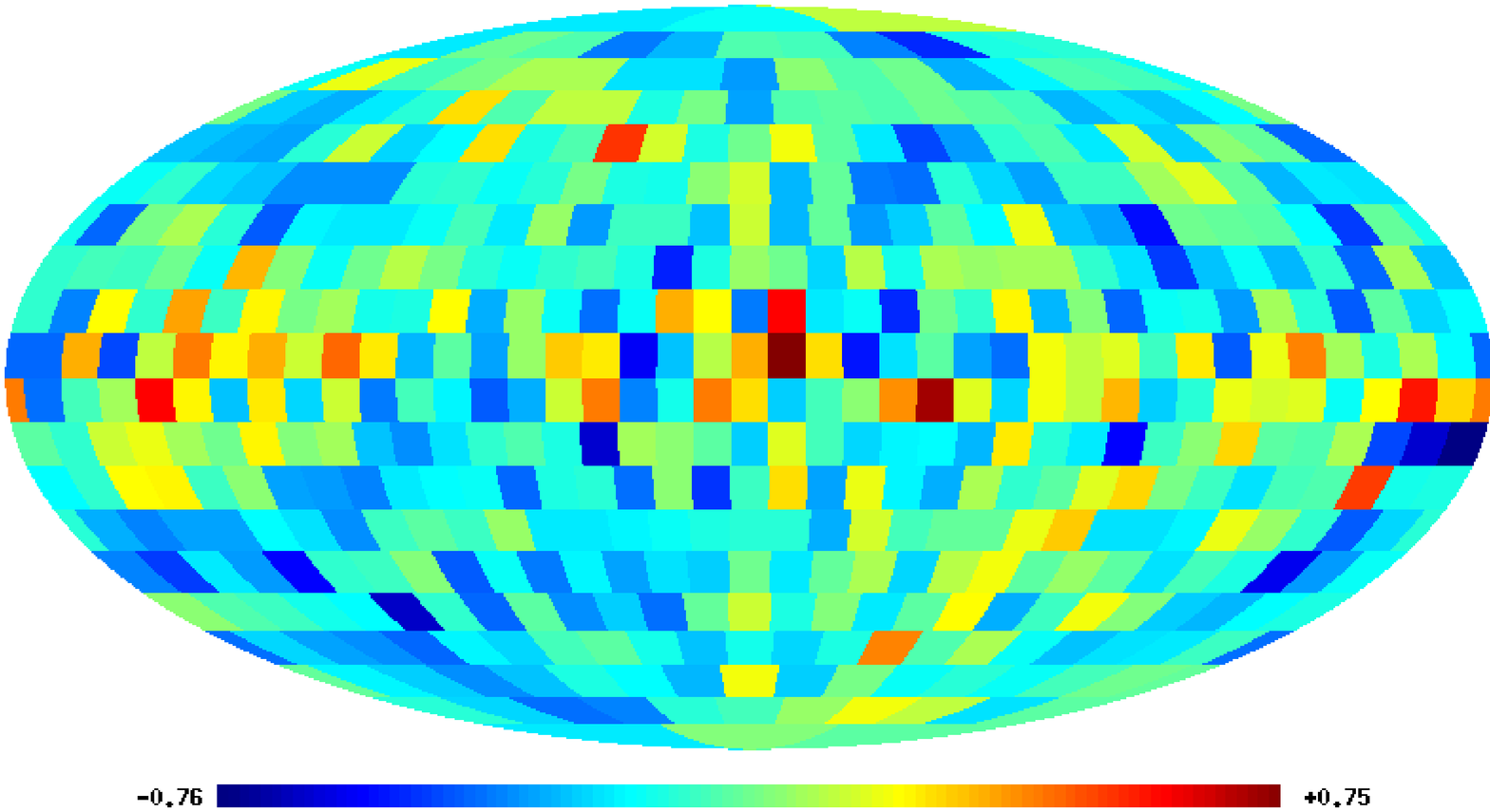,width=5cm,angle=-90}
\psfig{figure=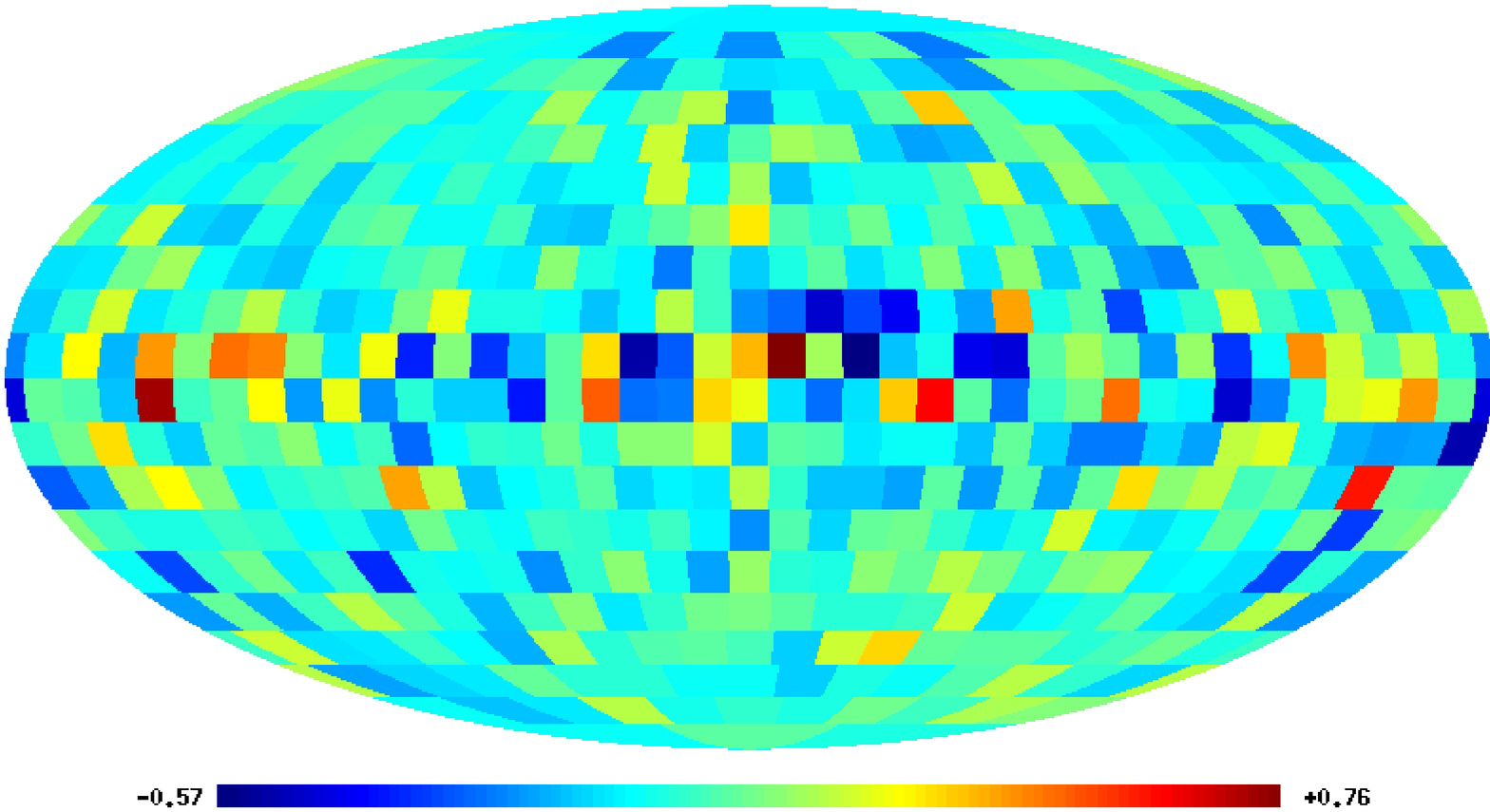,width=5cm,angle=-90}
\psfig{figure=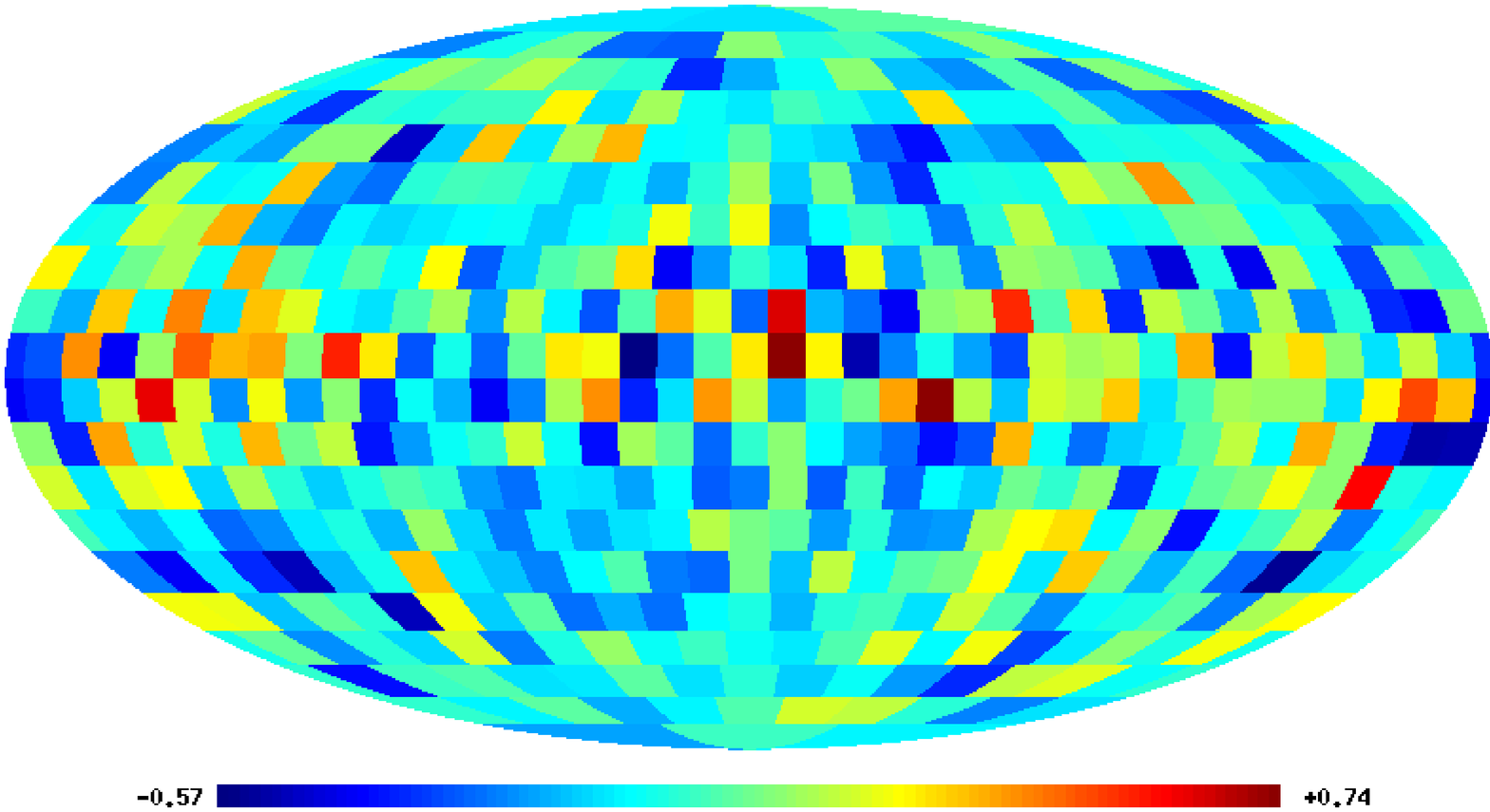,width=5cm,angle=-90}
} } }
\caption{ Correlation maps of ILC and
galactic foregrounds: dust, free-free and synchrotron radiation
(from left to right), for different correlation windows:
162\arcmin, 324\arcmin\ and 540\arcmin\ (from top to bottom). The
tone of the pixels reflects the value of the correlation
coefficients from --1.0 to 1.0. The brighter the tone the smaller
the absolute value of the correlation coefficient. }
\end{figure*}

We offer an approach realized within the framework of the GLESP
package \cite{glesp:Verkhodanov_n,glesp2:Verkhodanov_n}, which
allows detecting the correlations in the studied areas of a given
angular scale for the maps, pixelized with higher resolution. The
method is implemented in the pixel parametric space. The
correlation mapping consists of assigning the pixel number $p$
with a correlation result inside the solid angle $\Xi_p$,
calculated for two maps having higher resolution. As a result, we
obtain a new map, in which the value assigned to each pixel
reflects the correlation level of the studied maps in a given
area.

The correlation coefficient of anisotropy of the CMB temperature
and a given signal for every pixel $p$ ($p=1,2,...,N_0$, where
$N_0$ is the total number of pixels on the sphere), corresponding
to the solid angle $\Xi_p$ and deduced for the maps on the sphere
with an initial resolution, determined $\ell_{max}$ (hereafter the
value of $\ell_{max}=150$ is taken for the original maps), there
are {\small
\begin{eqnarray}
 K(\Xi_p|\ell_{max}) = \quad \quad \quad \quad \quad \quad \quad \quad \nonumber \\ \frac{\sum\sum\limits_{p_{ij}\in\Xi_p}
           (\Delta T(\theta_i,\phi_j) - \overline{\Delta T(\Xi_p))}
           (S(\theta_i,\phi_j) - \overline{S(\Xi_p))}}
        {\sigma_{\Delta T_p}\sigma_{S_p}}\,,
\end{eqnarray}}

\noindent where $\Delta T(\theta_i,\phi_j)$ is the anisotropy
value of the CMB temperature in a pixel with the coordinates
$(\theta_i,\phi_j)$ for the given pixelization resolution of the
sphere, \mbox{$S(\theta_i,\phi_j)$ is the value} of another signal
in the same area, $\overline{\Delta T(\Xi_p)}$ and
$\overline{S(\Xi_p)}$ are common values in the area $\Xi_p$,
deduced on the data from the maps with a higher resolution,
plotted by $\ell_{max}$, $\sigma_{\Delta T_p}$ and
\mbox{$\sigma_{S_p}$} are the corresponding standards for this
area.

\section{CORRELATION MAPS OF ILC WMAP5 AND FOREGROUND COMPONENTS}

One of the applications of the given approach is checking the
cleaning quality of the CMB signal isolated from multifrequency
observations. The presence of a residual (correlated) signal in
the CMB data may lead to a bias in power estimation in a number of
multipole ranges \cite{instab:Verkhodanov_n}, which in its turn
leads to a decrease in accuracy while determining the cosmological
parameters.

Using the WMAP data---a map of synchrotron radiation in the
K-band, a map of free-free emission in the V-band, dust radiation
in the W-band and, finally, the ILC map---we have built
correlation maps of foreground radiations and ILC at different
angular scales.

As it was demonstrated earlier in the literature dedicated to the
analysis of the ILC map and its signal statistics
\cite{ndv03:Verkhodanov_n,ndv04:Verkhodanov_n,nong:Verkhodanov_n},
there are some serious arguments for the fact that in this map at
different angular scales there exists some residual contribution
of foreground components, which, among other factors, gives the
found non-Gaussianity. This contribution may as well manifest
itself in the earlier discovered connection in the quadrupole
between a cleaned map of the microwave background and galactic
radiation components
\cite{instab:Verkhodanov_n,pecquad:Verkhodanov_n,nv_quad:Verkhodanov_n}.
According to the papers of the WMAP team
\cite{wmapresults:Verkhodanov_n,wmap5ytem:Verkhodanov_n}, the ILC
map is not intended for the CMB study on high multipoles, but it
could be used for a foreground components analysis. This is the
map we use for the foreground properties analysis.

In our previous work \cite{rzf_cmb:Verkhodanov_n}, using
one-dimensional sections, we employed a WMAP mission foreground
maps correlation search approach, and discovered in particular,
that in the WMAP maps sections on declination $\delta=41\degr$
there is a signal, correlated and anti-correlated with the data of
the components being separated. Among other things, on the angular
scales indicative for the Galactic plane (multipole range
$\ell$=10--20) some notable, compared to the simulated maps,
correlations with the galactic foreground components were
discovered. These correlations may be the consequence of the
residual errors at the signal separation phase.

In the present work we constructed the ILC and WMAP5 foreground
components correlation maps on the full sphere. We as well
estimated here the statistical value of the isolated correlated
areas using the Monte-Carlo method, comparing them with the
Gaussian random fields maps, generated on full celestial sphere
for the $\Lambda$CDM cosmological
model~\cite{wmap5ycos:Verkhodanov_n}. The results of correlation
mapping of the ILC and galactic foreground maps are demonstrated
in Fig.\,1 for the correlation windows with the sides of
162\arcmin, 324\arcmin, 540\arcmin\, (consequently, $\ell_{max_1}
=$ 33, 16, and 10) for the three phase components. The window
dimensions were chosen in such a manner that they fit into the
range of angular scales of the Galaxy's influence ($\ell$=10--20),
and the possible Sachs-Wolfe effect \mbox{($\ell\le45$).}

The Galactic plane can clearly be seen on the maps presented in
Fig.1. An important marker here is the correlation value
statistics for every pixel. We constructed the histograms for all
the given maps and compared them with analogous results for the
simulated maps. The simulated maps were generated with an
assumption that the CMB data are the result of a Gaussian random
process with an angular power spectrum, corresponding to the
cosmological \mbox{$\Lambda$CDM model.} In total, there were
constructed 50 random simulated maps using the {\it cl2map}
procedure of the GLESP \mbox{package
\cite{glesp:Verkhodanov_n,glesp2:Verkhodanov_n},} and with their
aid we were able to evaluate the admissible interval, determined
by the correlation values spread dispersion. The pixel value
distribution is presented in Fig.\,2. The admissible limits of the
correlation level constructed on the modelling data, are shown
with dashed lines.

\begin{figure*}[tbp]
\centerline{
\vbox{
\hbox{
\psfig{figure=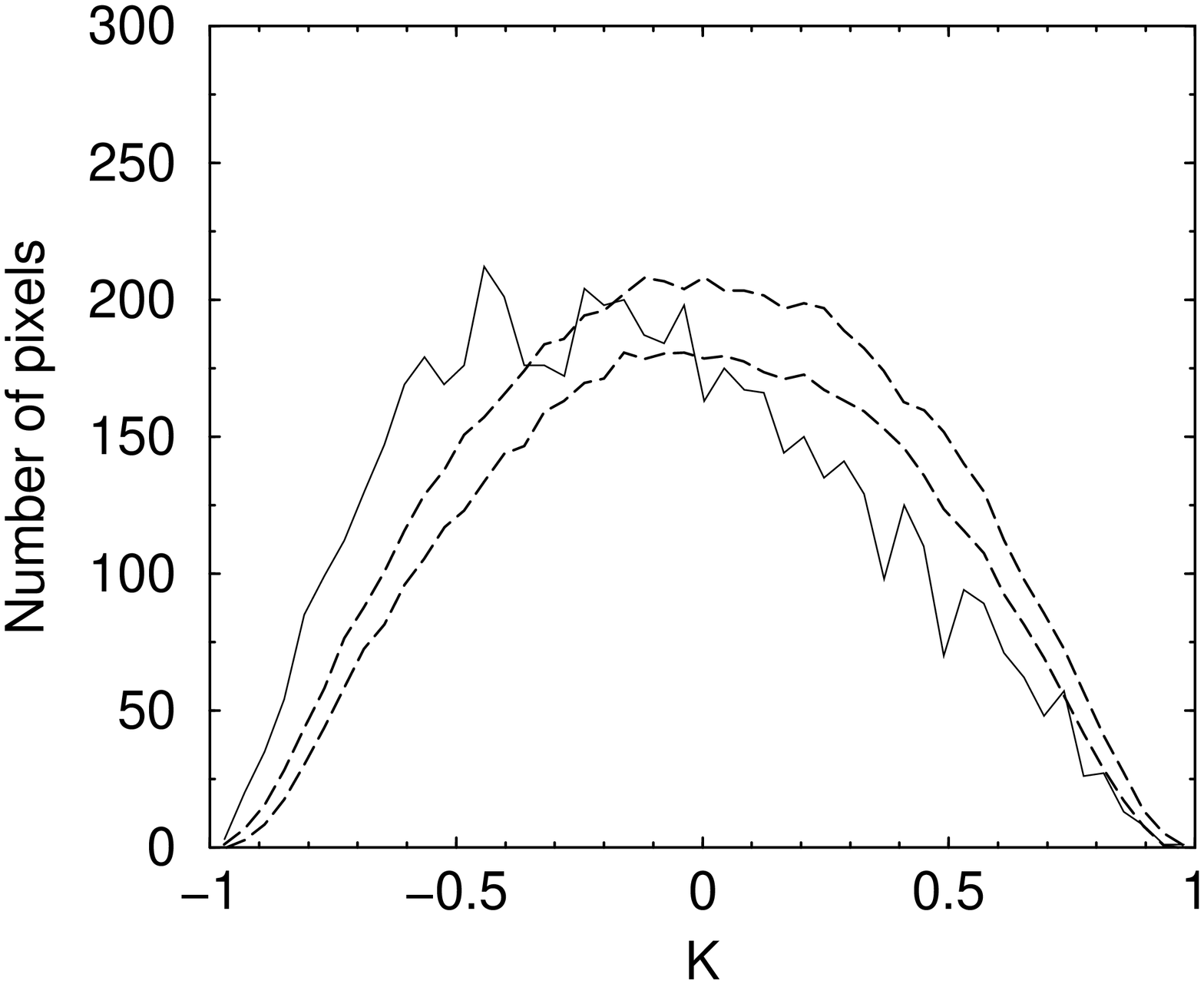,width=5cm,angle=-90}
\psfig{figure=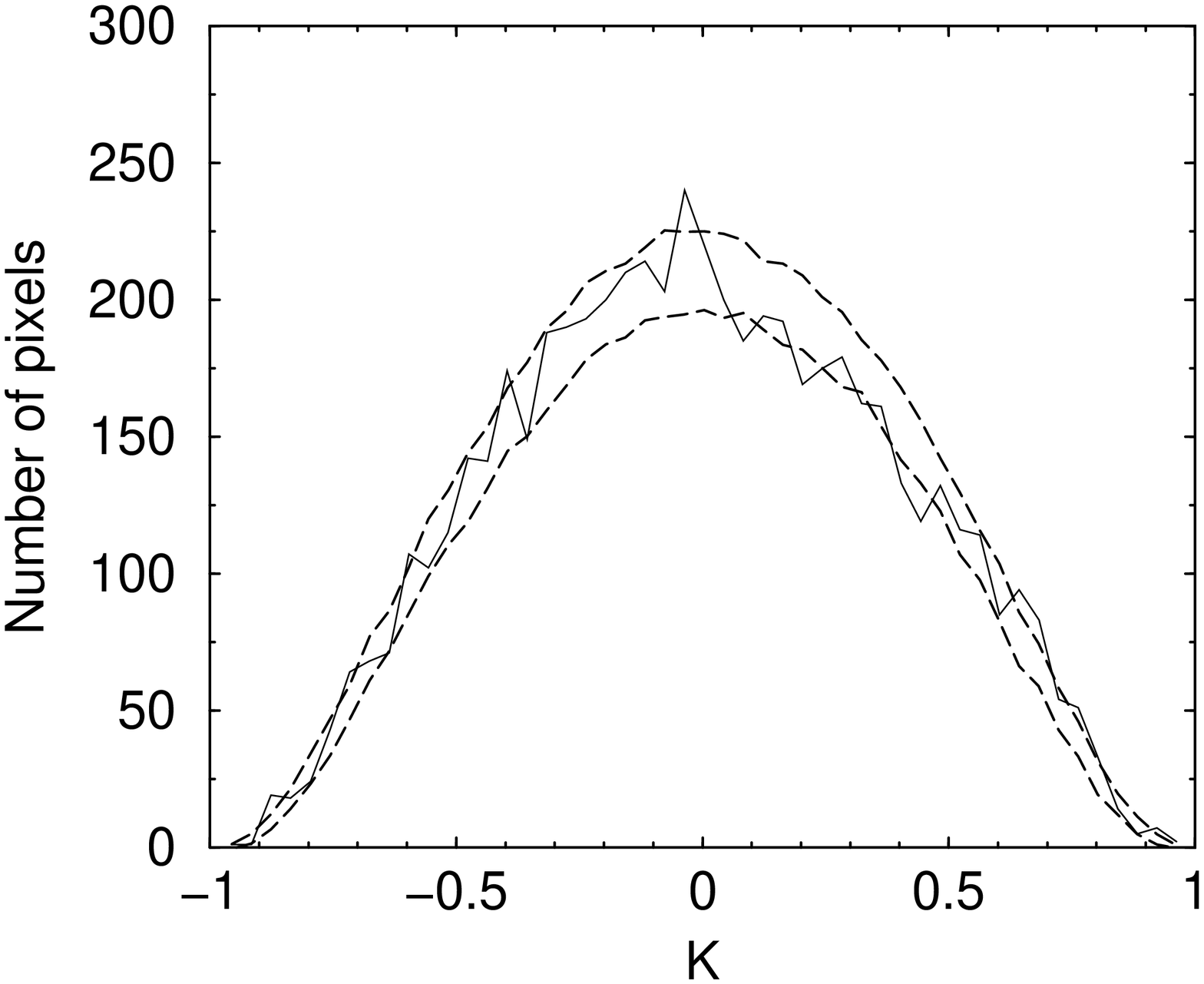,width=5cm,angle=-90}
\psfig{figure=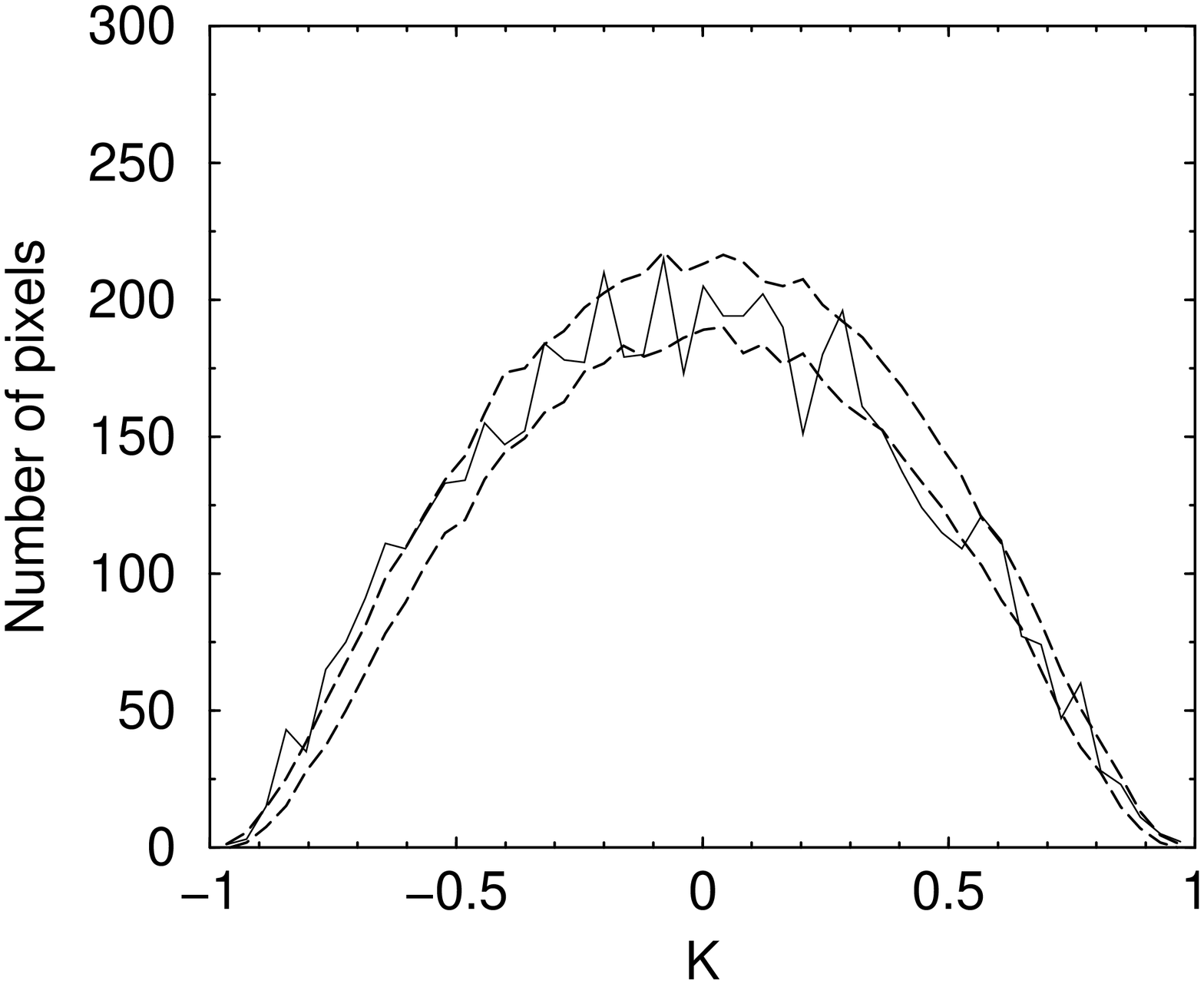,width=5cm,angle=-90}
}
\hbox{
\psfig{figure=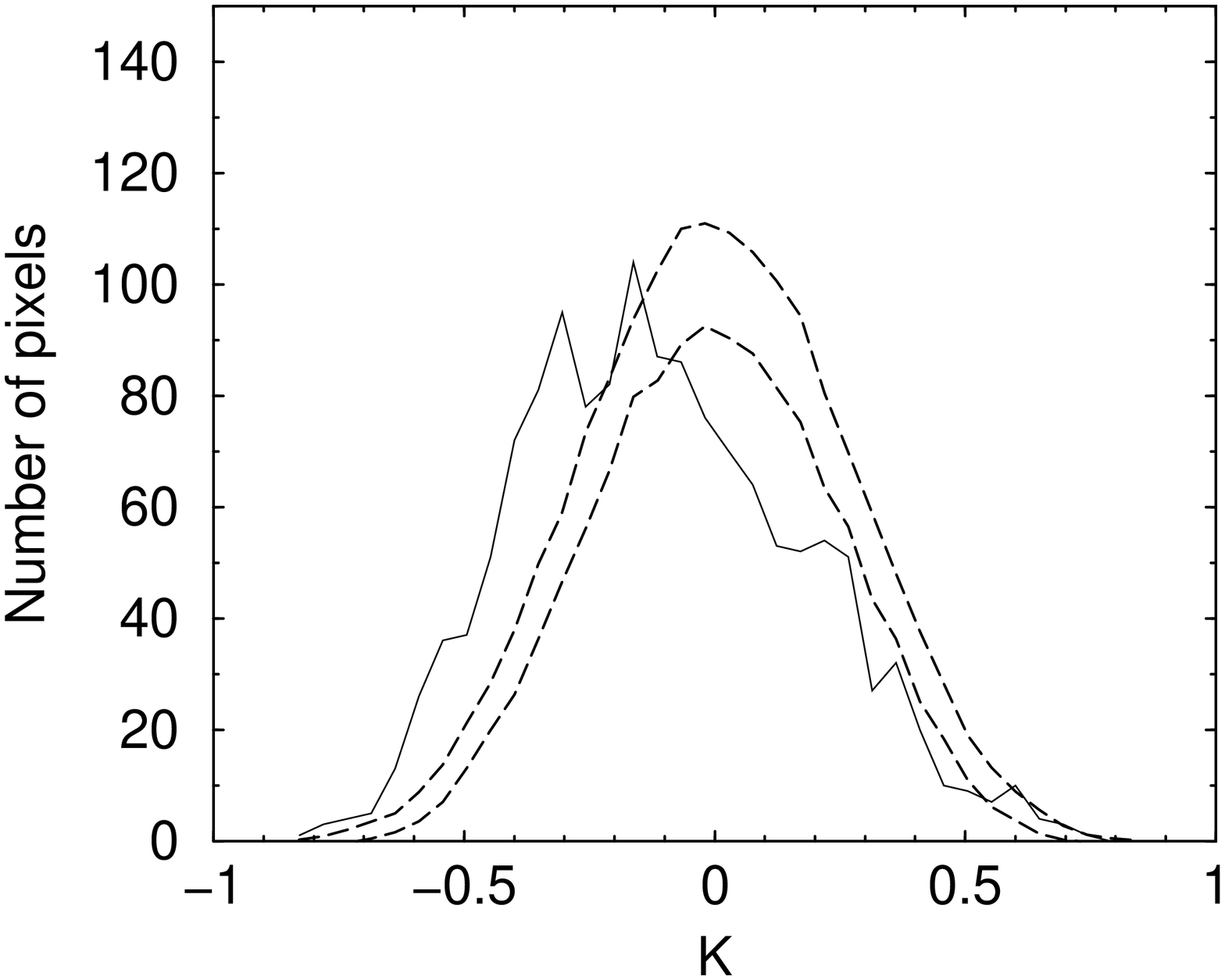,width=5cm,angle=-90}
\psfig{figure=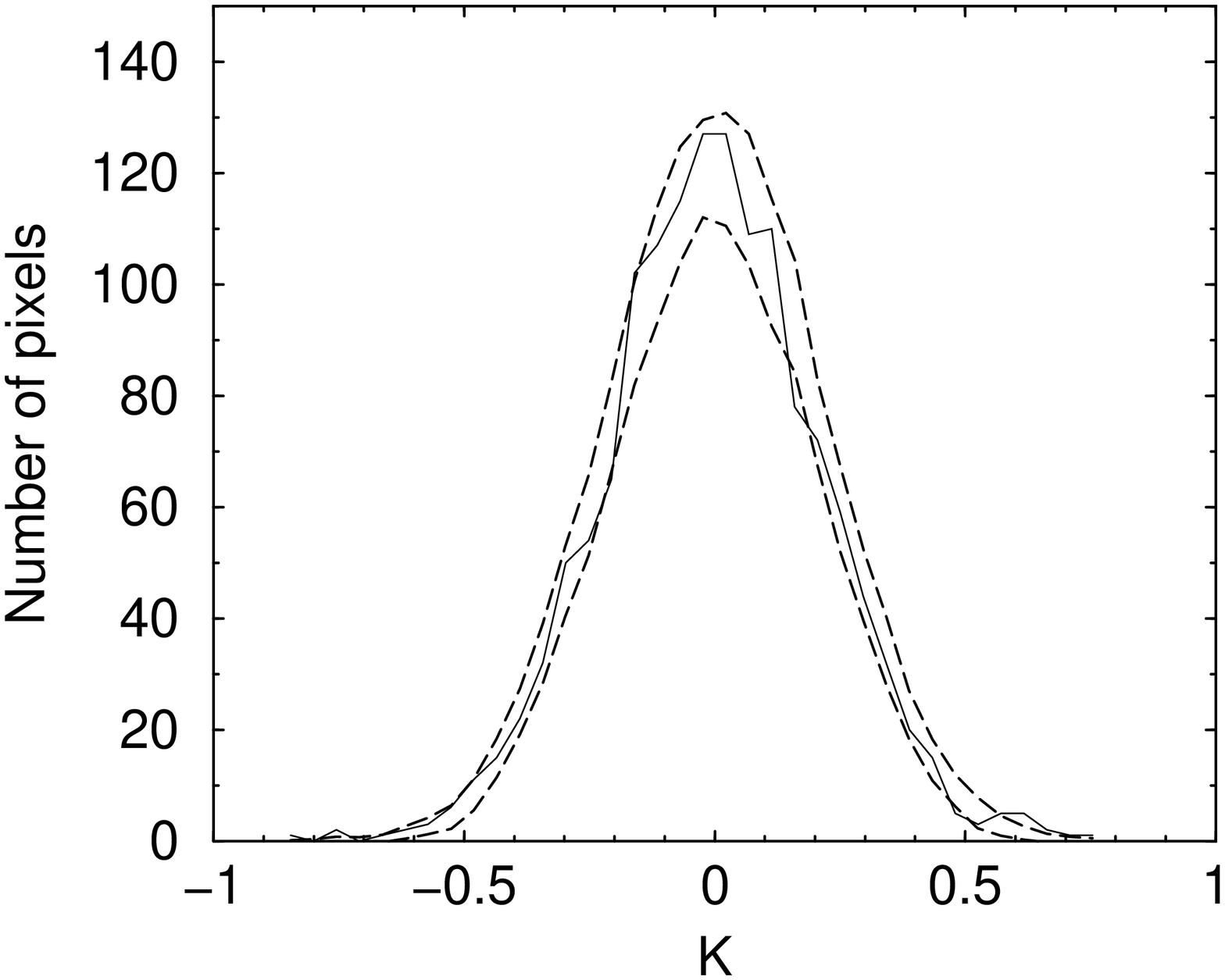,width=5cm,angle=-90}
\psfig{figure=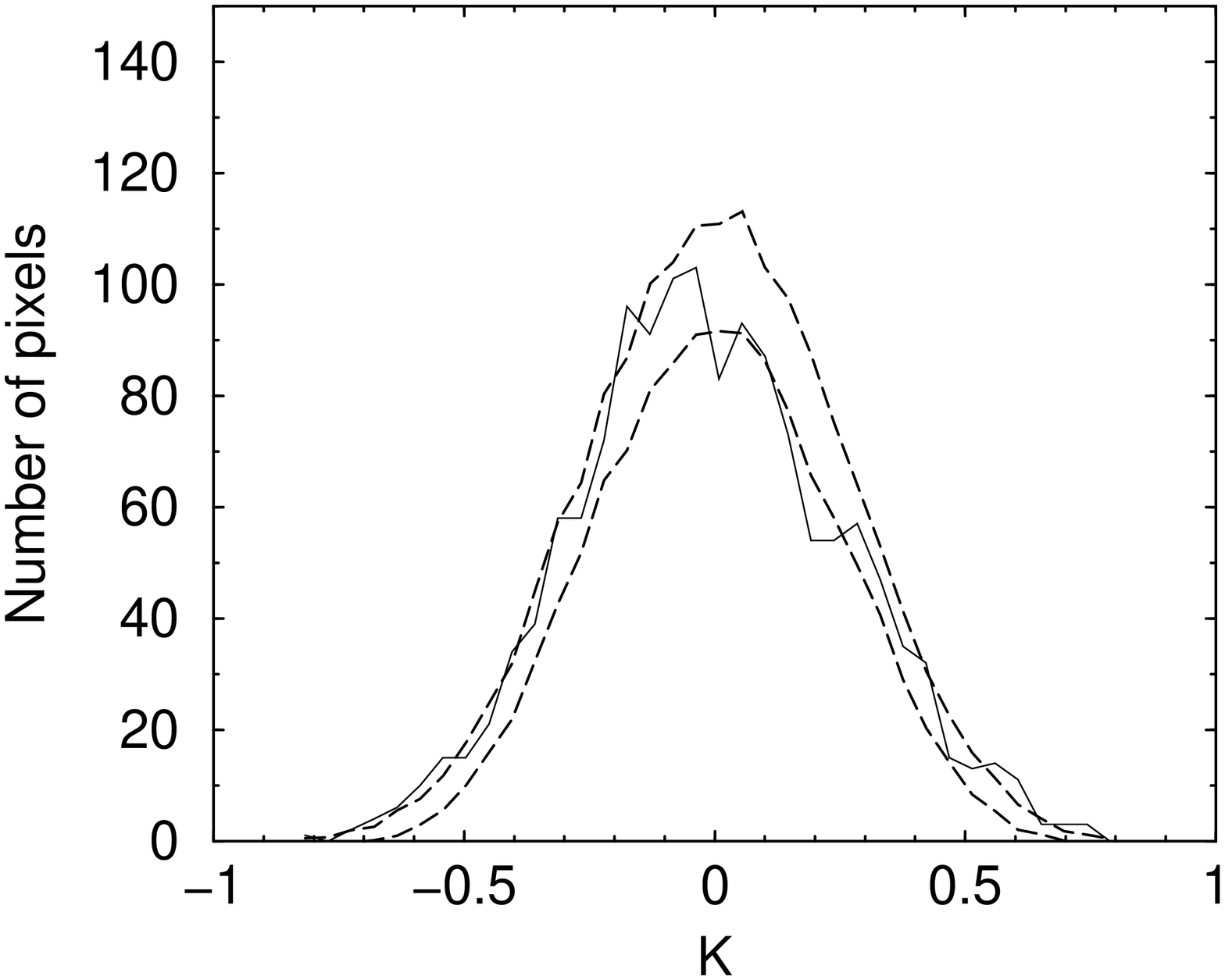,width=5cm,angle=-90}
}
\hbox{
\psfig{figure=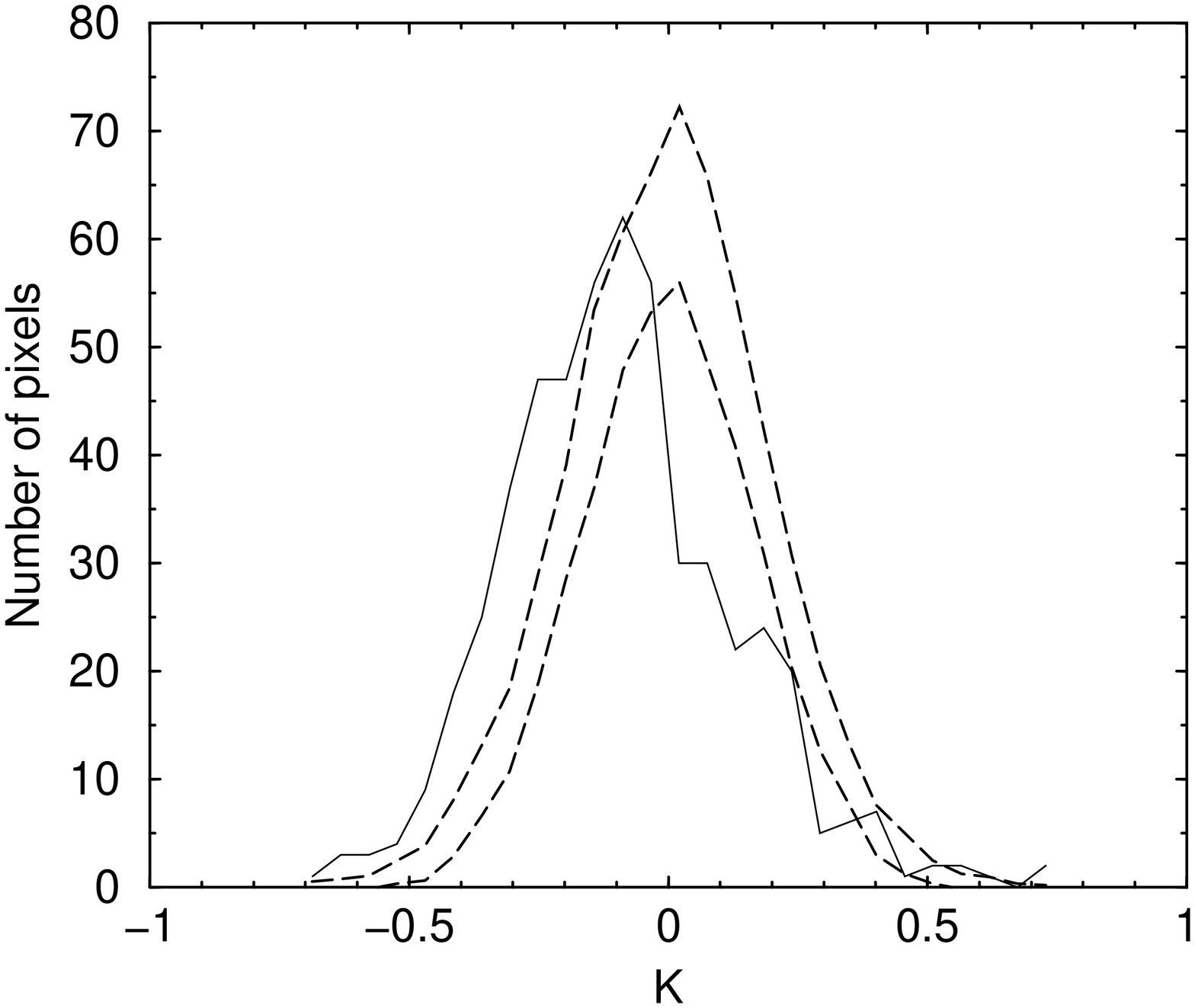,width=5cm,angle=-90}
\psfig{figure=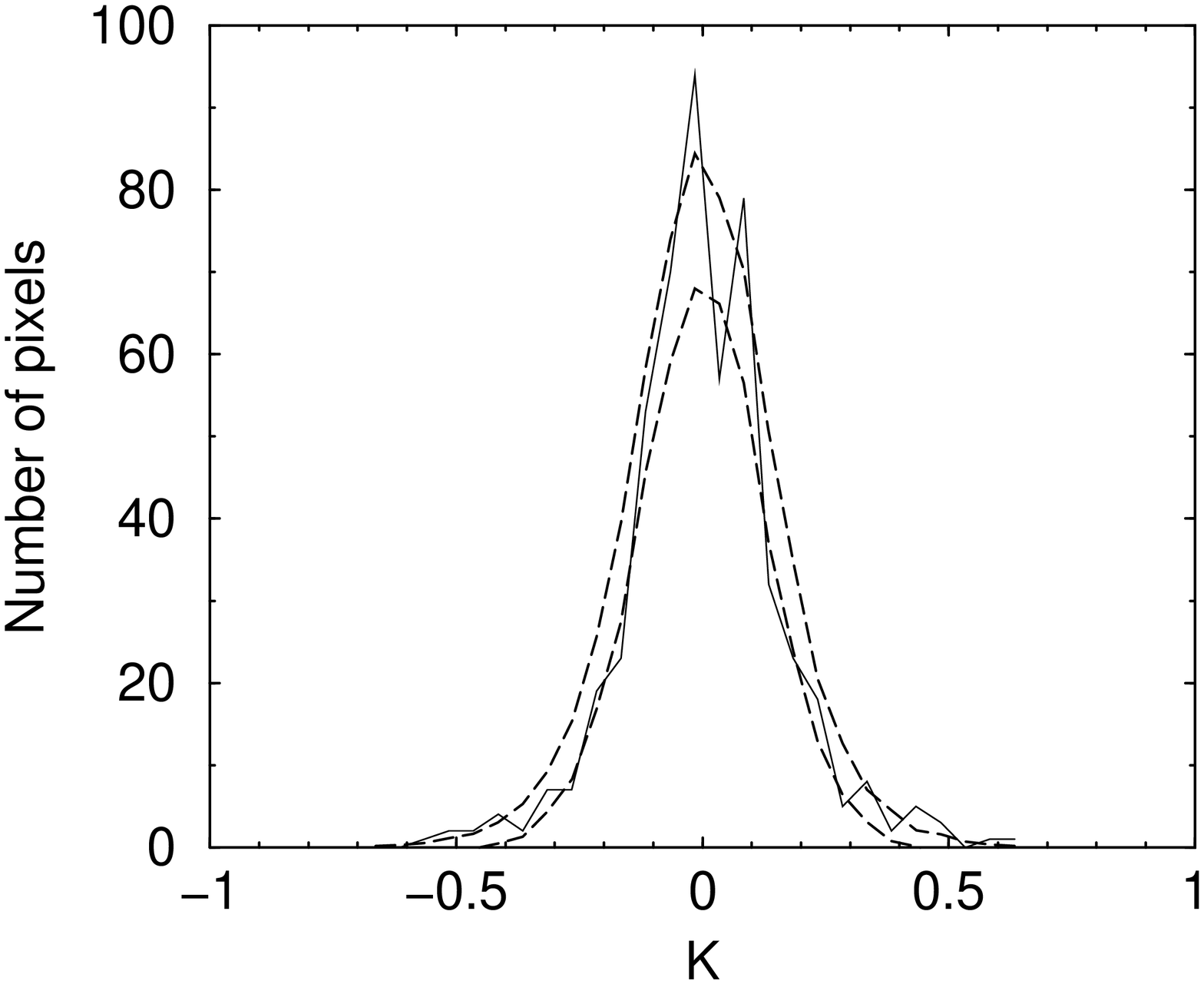,width=5cm,angle=-90}
\psfig{figure=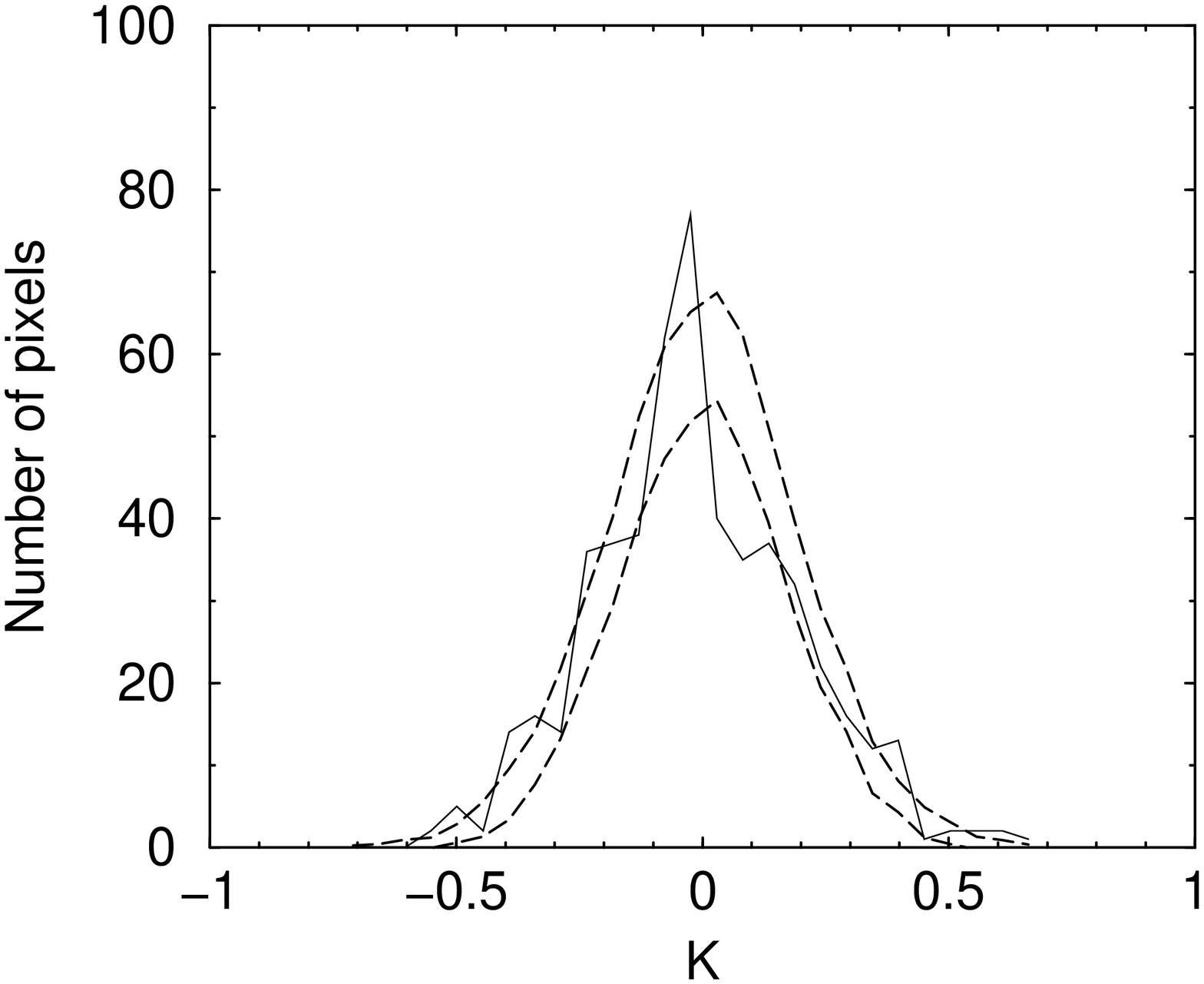,width=5cm,angle=-90}
} } }
\caption{Distribution of correlation
index values in the pixels of the ILC and galactic foregrounds
correlation maps: the dust radiation, the free-free and
synchrotron emission (from left to right) for different
correlation windows: 162\arcmin\,, 324\arcmin\, and 540\arcmin\,
(from top to bottom). The admissible limits of correlation level,
constructed on the modelling data, are shown with dashed lines. }
\end{figure*}

Two foreground components correlations, dust and synchrotron,
exhibit a shift in the range of anti-correlations. While in case
of the dust component this shift is observed in all the studied
scales, in case of synchrotron radiation it is presented on the
histograms for 324\arcmin~and 540\arcmin. The free-free emission
histogram shows the distribution within the normal range,
corresponding to CMB as a Gaussian random process. We have to
note, that an analogous shift in the distribution was detected for
low multipoles at the analysis of the ILC signal reconstruction
\mbox{method \cite{instab:Verkhodanov_n,nv_quad:Verkhodanov_n}}
via regenerating the simulated maps with a further intermixing
with the foreground components, reconstructing the signal, then
correlating the reconstructed maps and constructing the
correlation coefficients distribution. In this approach the
distribution evaluation consisted of two steps only: the
correlation map construction and the calculation of the pixel
values distribution histogram.

A shift in the correlation coefficient distribution may be linked
to an underestimation in the model used of the dust distribution
in our Galaxy, where the excess in the negative coefficient values
supports the hypothesis that in the corresponding pixels the
correlated CMB and dust signals have an opposite sign. This fact
can be used for the correction of dust component contribution. We
are planning further studies of the detected bias.

\section{ILC WMAP5 AND NVSS CORRELATION MAPS}

Let us consider another correlation mapping application. Among the
CMB map research works there are those dedicated to the dark
matter studies using the Sachs-Wolfe integral effect
\cite{swe:Verkhodanov_n}, registered while correlating the CMB
maps and the NVSS source distribution
\cite{nolta_cmb_de:Verkhodanov_n,mcewen_cmb_de:Verkhodanov_n}.

The NVSS survey \cite{nvss:Verkhodanov_n} is the most
comprehensive Northern sky survey. It was conducted using the NRAO
Very Large Array telescope (VLA) at a frequency of 1.4 GHz from
1993 to 1996, it has high sensitivity and covers the sky north of
the declination $\delta=-40\degr$ (33884\sqd~or 82\% of the
celestial sphere). This survey is actively used for various
statistical studies in cosmology. The catalogue of this survey
contains $1.8\times10^6$ sources and, according to it description,
it is 99\% complete up to the integrated flux densities over
$S_{1.4\,\mbox{\tiny\rm GHz}}=3.5$\,mJy and 50\% complete up to
the flux densities of 2.5\,mJy. The survey used the
D--configuration of the VLA radiotelescope, and the size of the
synthesized direction diagram in the half-power level, determining
the resolution, constituted approximately $45''$. The survey data
are available from the NRAO website\footnote{\tt
http://www.cv.nrao.edu/nvss/}, the virtual telescope
SkyView\footnote{\tt http://skyview.gsfc.nasa.gov} and from the
CATS database\footnote{\tt http://cats.sao.ru}
\cite{cats:Verkhodanov_n,cats2:Verkhodanov_n}.

Some non-Gaussian peculiarities were detected from the WMAP and
NVSS data, poorly matching with the inflation $\Lambda$CDM model.
Among them, for example, is the Cold Spot (CS), having the size of
about 10\degr~ and the galactic coordinates of ($l$=209\degr,
\mbox{$b$=--57\degr).} The CS, being statistically isolated in the
CMB distribution, is one of the maps' particularities that
contradict the hypothesis of homogeneous Gaussian background
fluctuations. Originally, it was noted as a deviation from the
Gaussian statistics while using the \mbox{wavelet analysis
\cite{vielva04:Verkhodanov_n,cruz05:Verkhodanov_n}} for the first
year data recorded by the WMAP mission. Subsequently, in the NVSS
maps there was detected a significant dip of the radio source
density of matter in the CS region \cite{rudnick:Verkhodanov_n},
that allowed to hypothesize the existence of a Great Void 140\,Mpc
across, at redshift of $z<1$, causing a gravitational anomaly
leading to the integrated Sachs-Wolfe effect
\cite{swe:Verkhodanov_n}, and manifesting itself as a CS. At the
same time there appeared some serious facts arguing that in this
map at different angular scales there is some residual foreground
components contribution, which produces the detected
non-Gaussianity. This contribution may manifest itself in the
earlier discovered link in the quadrupole between the cleaned CMB
map and the galactic radiation components
\cite{pecquad:Verkhodanov_n,nv_quad:Verkhodanov_n} and, for
another hand, determine the low multipole properties $\ell\le20$,
resulting in their unstable reconstruction
\cite{instab:Verkhodanov_n,nv_quad:Verkhodanov_n}. Specifically,
the CS particularities may be explained via the effects of these
multipoles: the deviations of peak clusters statistics around the
spot \cite{nas_cs:Verkhodanov_n}, namely, an increase in the the
positive peaks number. An independent study of the properties of a
spot, detected in the region with close coordinates on the maps
and the radio source count in the NVSS survey
\cite{nvss:Verkhodanov_n}, has demonstrated that the studied Cold
Spot, with its gigantic size and its mere existence being
difficult to explain in terms of the existing cosmological
$\Lambda$CDM model, may as well be a simple statistical deviation,
caused by the systematic \mbox{effects
\cite{no_cs:Verkhodanov_n}.}

\begin{figure*}
\centerline{
\hbox{
\psfig{figure=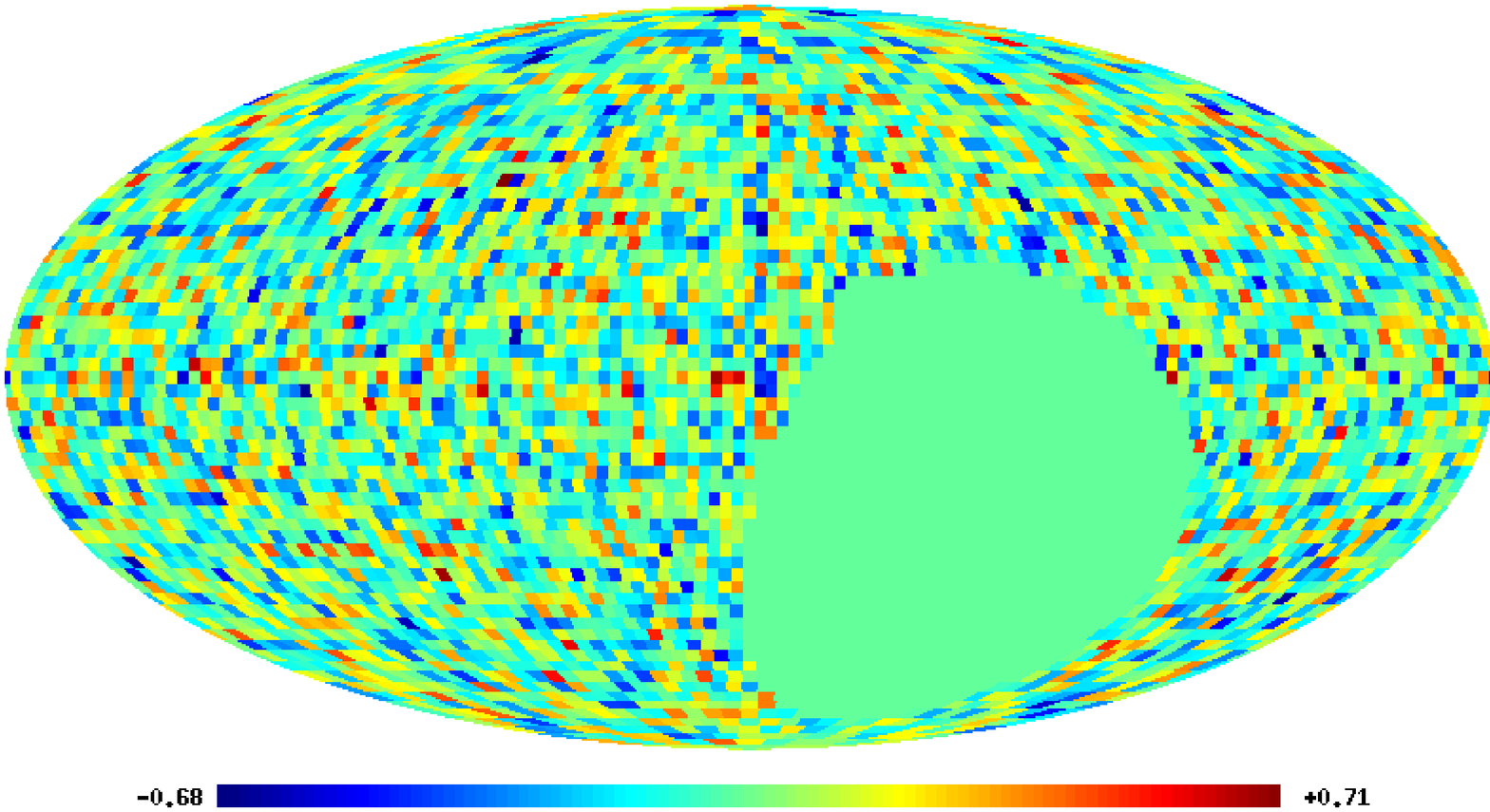,width=5cm,angle=-90}
\psfig{figure=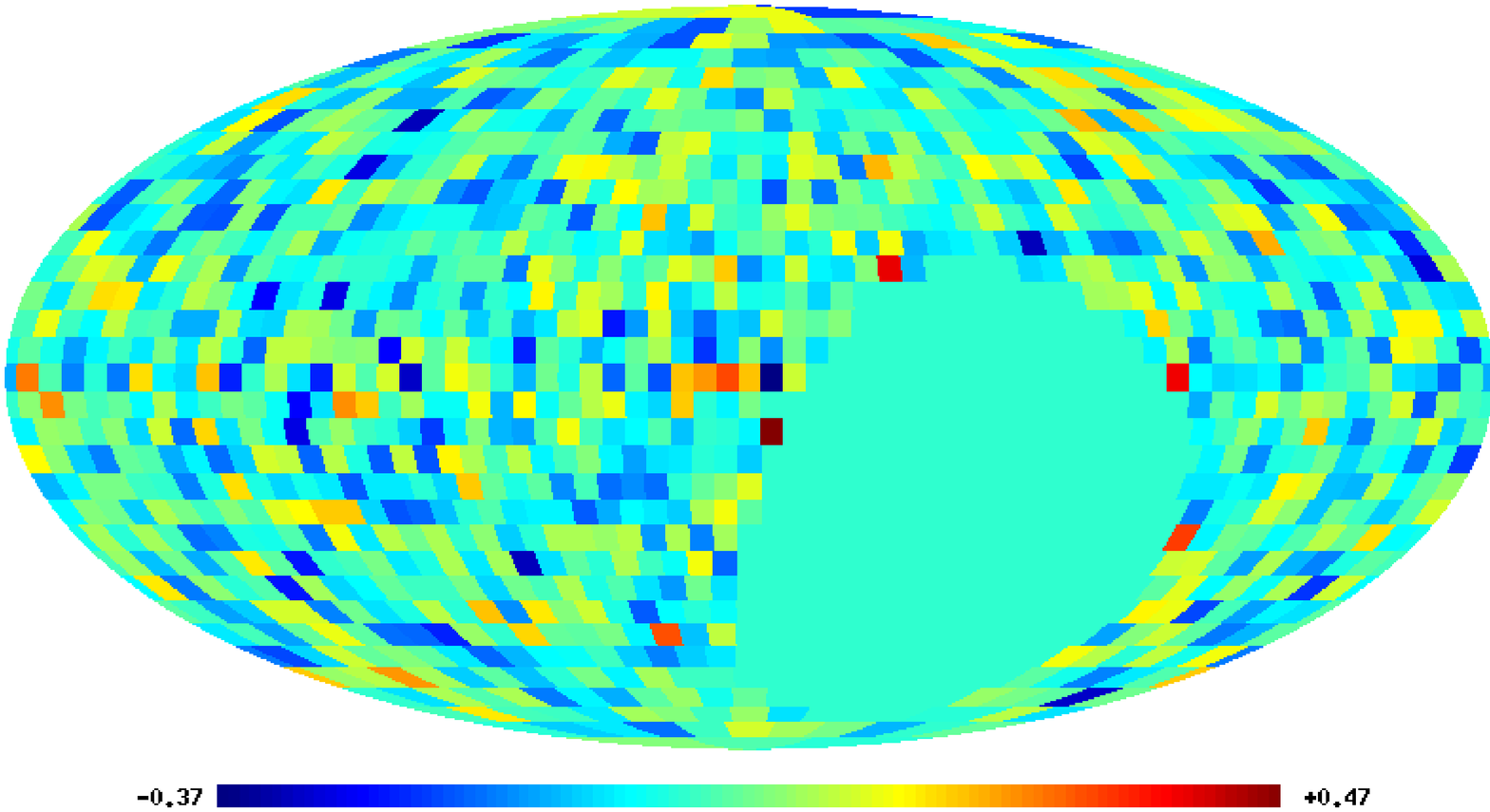,width=5cm,angle=-90}
\psfig{figure=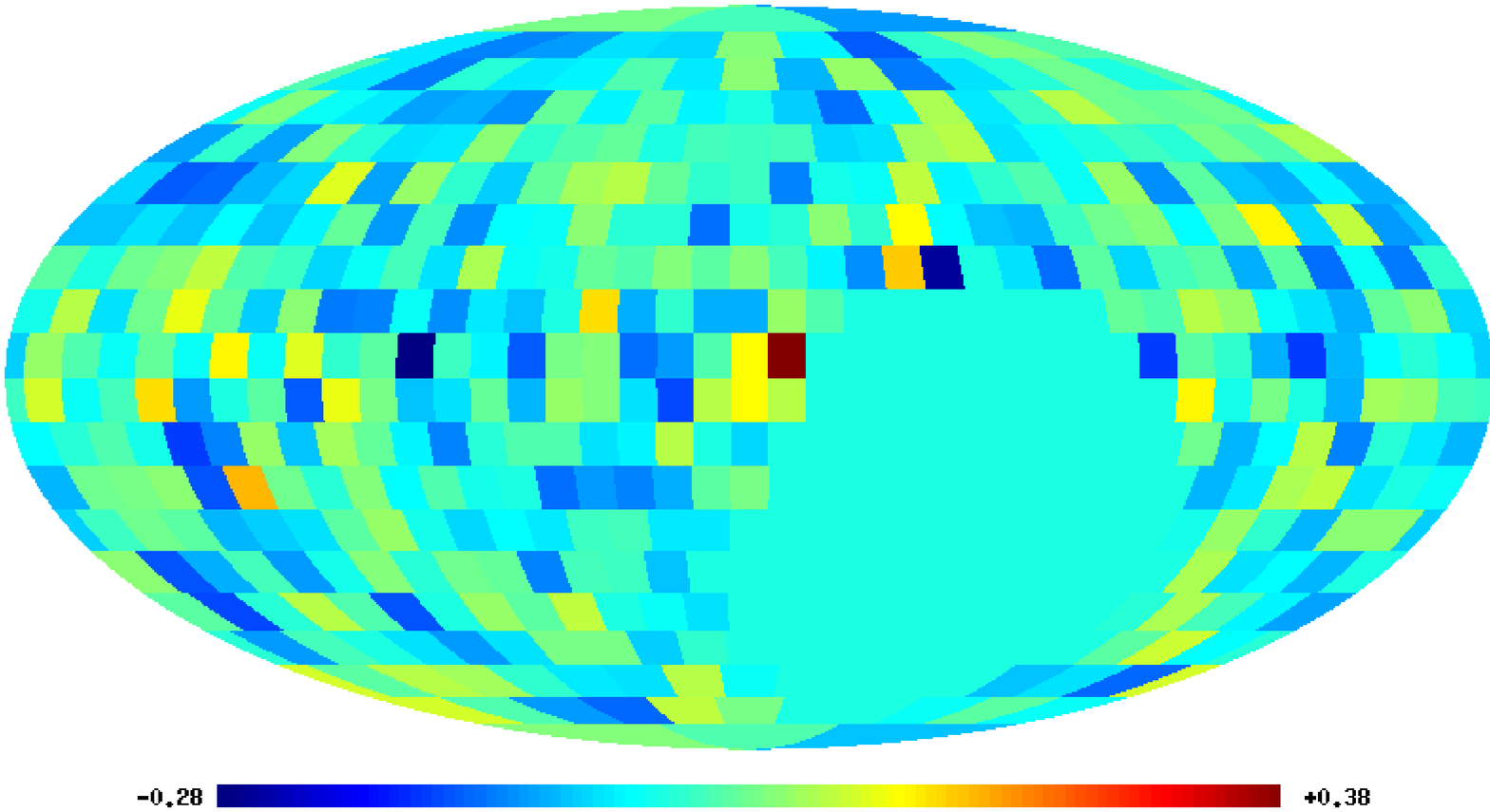,width=5cm,angle=-90}
} }
\caption{ ILC and NVSS correlation maps
for different correlation windows: 162\arcmin\,, 324\arcmin\ and
540\arcmin\, (from right to left). In the right bottom part of
each figure there is an screened area, which was not considered in
our analysis due to the lack of NVSS observations in the
$\delta<-40\degr$ range. }
\end{figure*}

\begin{figure*}
\centerline{
\hbox{
\psfig{figure=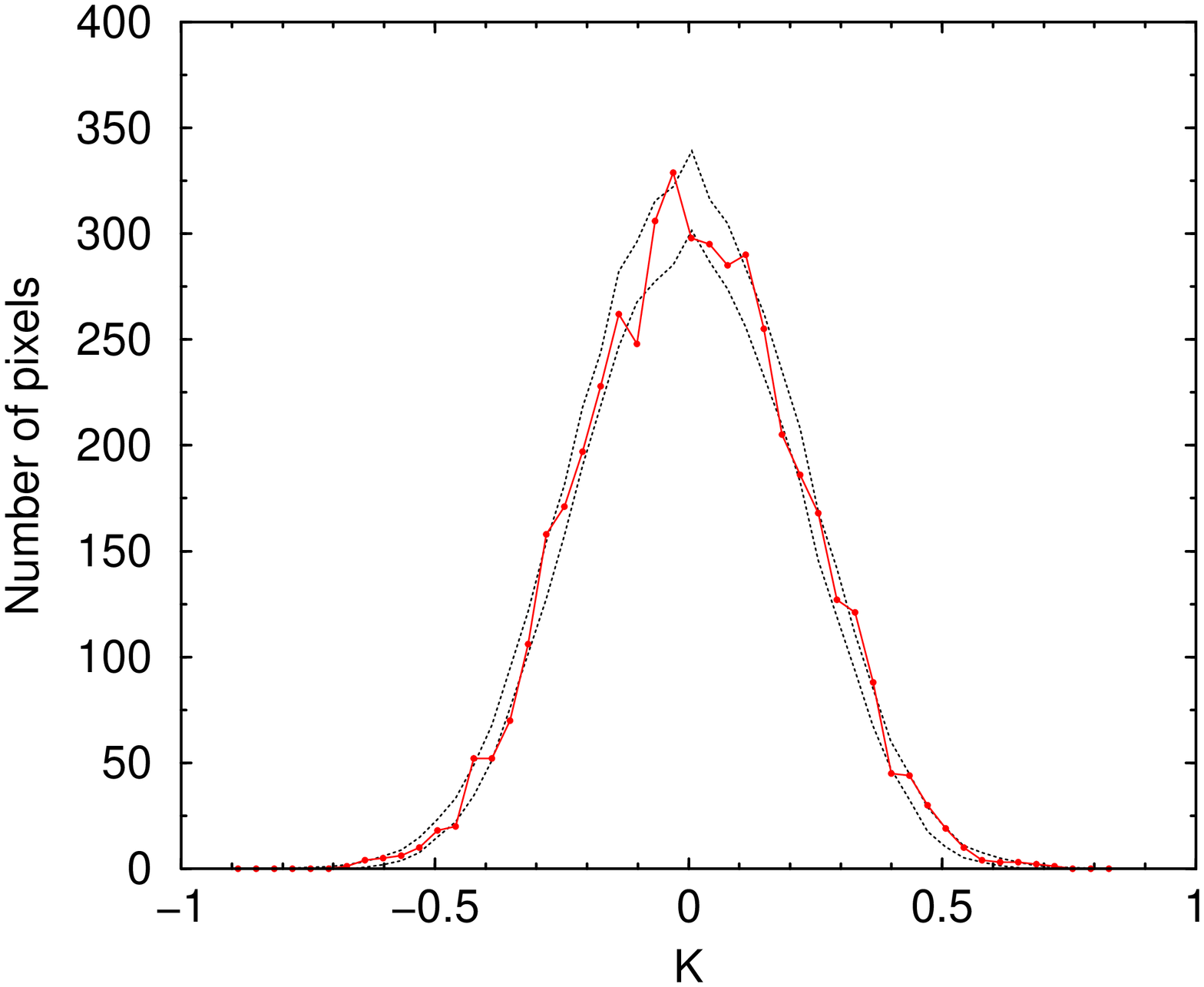,width=5cm,angle=-90}
\psfig{figure=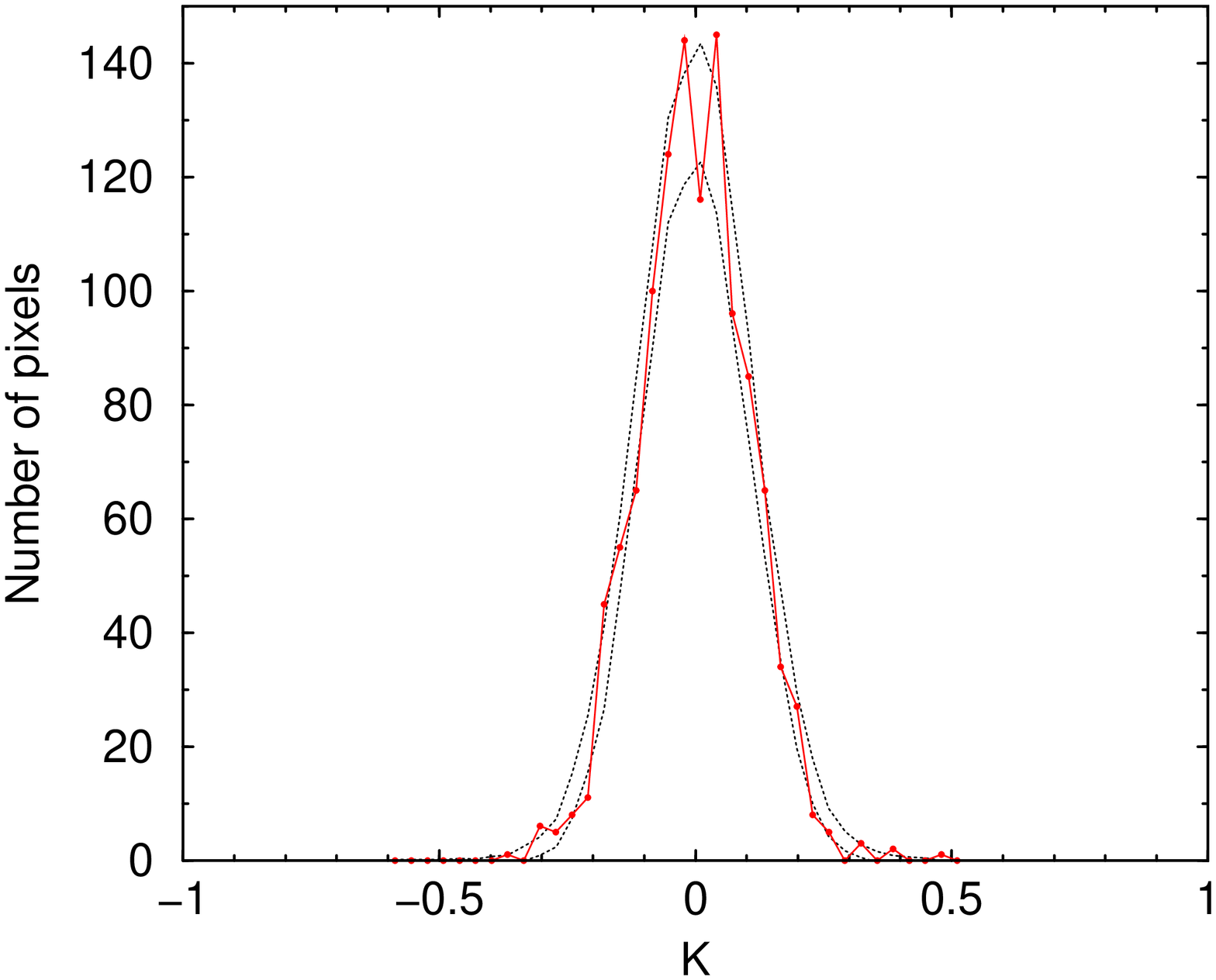,width=5cm,angle=-90}
\psfig{figure=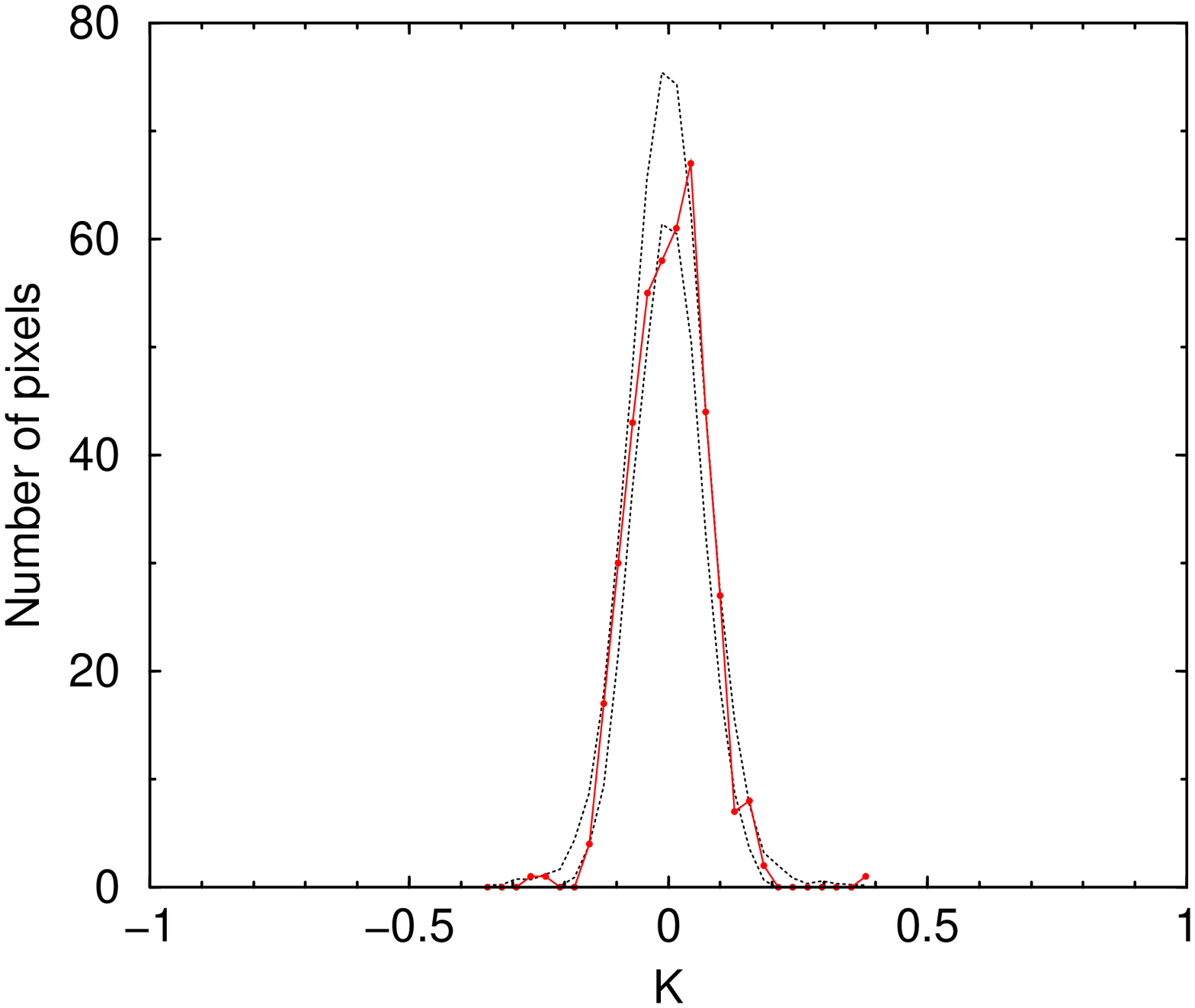,width=5cm,angle=-90}
} }
\caption{ Distribution of correlation
index values in the pixels of the ILC and NVSS correlation maps
for different correlation windows: 162\arcmin\,, 324\arcmin\,, and
540\arcmin~(from right to left). The admissible limits of the
correlation level, constructed using the modelling data, are shown
by the dashed lines. }
\end{figure*}

Earlier we made a correlation analysis of the NVSS and WMAP5
surveys, preliminarily modifying the NVSS map making use of the
method of one-dimensional sections for different angular scales
\cite{nvss_cmb:Verkhodanov_n}. To construct the modified NVSS map,
we used a characteristic, represented by a mean square of the
source flux density in an area of given dimensions with the center
at the pixel, determined by the selected area. As a result we
discovered that the correlation statistics in the studied scales
(0.75\degr, 3\degr, 4.5\degr, and 6.75\degr) does not vary from
the expected random maps. Therefore, we may conclude that the
found correlations may be caused by a statistical coincidence.

In the present work we carried out the mapping of the correlation
properties on the sphere using the same data from WMAP5 and NVSS.
The results are shown on Fig.\,3 and 4. While analyzing the map
statistics (Fig.\,4) we were taking into account the absence of
NVSS data in the declination area of \mbox{$\delta<-40\degr$}. The
NVSS map was constructed using two methods: (1) via computing the
number of sources in every pixel and (2) via summing up the flux
densities of the radio sources, contained in the pixel area. As
the correlation properties are practically undistinguishable (the
difference of less than 1\%), we shall demonstrate the whole
analysis for the first case.

As we can see from Fig.\,4, the statistics of the correlated pixel
values lies within the admissible random deviations, what confirms
the earlier \mbox{estimations \cite{nvss_cmb:Verkhodanov_n}.} This
implies that in the correlation coefficient distribution no signal
is found with a level higher than 1$\sigma$ on the given scales,
caused by, for example, an interaction with the large-scale
structure and dark energy, manifesting themselves through the
Sachs-Wolfe effect. On the other hand, as the coefficient
distribution corresponds to the expected random deviations for the
Gaussian disturbances in the $\Lambda$CDM model, we may arrive at
a conclusion that there exists an infinitesimal number of
separated non-Gaussian spots, if any at all. And allowing for the
fact that the Cold Spot itself might be the result of a local
modulation of spherical harmonics, that appeared as a consequence
of the component separation \cite{nas_cs:Verkhodanov_n}, we may
construe that this phenomenon, observed simultaneously in CMB and
NVSS, is most likely not an anomaly but may be a systematism
(foreground overestimation) while analysing the WMAP data and a
random coincidence with a statistically insignificant peculiarity
of the NVSS distribution.

\section{CONCLUSION}

We have presented here a method of correlation mapping on the
sphere for a given angular scale. This method is
software-implemented within the framework of the GLESP package.
The proposed method allows analyzing on the sphere the properties
of a random CMB signal having a single realisation based on its
statistical characteristics only, namely, on its ergodicity, when
taking a multitude of CMB realisations in different regions of the
sphere, we can make a conclusion about its realisation in a
multitude of similar Universes, and, in so doing, evaluate its
probable values. We have demonstrated the capabilities of our
method for the WMAP5 data (the maps of ILC and foreground
radiations: synchrotron, free-free and dust). For the correlation
maps constructed, we discovered a shift in the pixel value
distribution in favour of anti-correlations for dust and
synchrotron radiation on the scales of 162\arcmin, 324\arcmin, and
540\arcmin. This calls for a possible overestimation of the
contribution of these foregrounds in the method of ILC components
separation. An excess in the negative coefficient values indicates
that in the corresponding pixels, the correlated signals of CMB
and dust have an opposite sign. This fact may be used while
correcting the contribution of the dust component. We are planning
further studies of this issue.

The described approach was as well used for a study of correlation
properties of the CMB and NVSS maps. In the correlation
coefficient distribution there was found no additional signal a
level higher than 1$\sigma$ over the one expected in the event of
a random agreement of Gaussian disturbances in the $\Lambda$CDM
cosmological model. Such a distribution of correlations argues for
the fact that the Cold Spot, observed both in the CMB and NVSS,
may be in fact the result of a random coincidence, and there is no
need invoking new physics to explain it.

\noindent
{\small
{\bf Acknowledgments}.
We would like to thank NASA for the chance to use its Legacy
Archive for Microwave Background Data Analysis from which we
retrieved the WMAP data. We are as well grateful to the authors of
the \linebreak HEALPix\footnote{{\tt
http://www.eso.org/science/healpix/}} \cite{healpix:Verkhodanov_n}
package using which we converted the WMAP maps into the $a_{\ell
m}$ coefficients. In our work we used the GLESP\footnote{{\tt
http://www.glesp.nbi.dk}}
\cite{glesp:Verkhodanov_n,glesp2:Verkhodanov_n} package for the
further  analysis of the CMB data on the sphere and the
FADPS\footnote{{\tt http://sed.sao.ru/$\sim$vo/fadps\_e.html}}
\cite{fadps:Verkhodanov_n,fadps2:Verkhodanov_n} one-dimensional
data processing system. The present work was supported by the
Leading Scientific Schools of Russia grant and the Russian
Foundation for Basic Research (project Nos. 09-02-00298 and
09-02-92659-IND). O.V.V. as well acknowledges the partial support
of the RFBR (project No. 08-02-00159).
}

\end{document}